\UseRawInputEncoding
\documentclass[referee]{raa}

\usepackage{graphicx,times}
\usepackage{natbib}
\usepackage{amssymb,amsmath}
\usepackage{color}
\bibpunct{(}{)}{;}{a}{}{,}

\usepackage[a4paper=true,pagebackref=true]{hyperref}
\hypersetup{colorlinks = true, linkcolor = green, anchorcolor = red, citecolor = blue, filecolor = red, pagecolor = red, urlcolor = red}
\begin{document}

   \title{Gamma-Ray Bursts with Extended Emission$:$ Classifications, Energy Correlations and Radiation Properties
}

 \volnopage{ {\bf 20XX} Vol.\ {\bf X} No. {\bf XX}, 000--000}
   \setcounter{page}{1}

   \author{X. L. Zhang\inst{1\dag}, C. T. Zhang\inst{1\dag}, X. J. Li\inst{1}, F. F. Su\inst{1,3}, X. F. Dong\inst{1}, H. Y. Chang\inst{2}, Z. B. Zhang\inst{1\ddag}}
 \renewcommand{\thefootnote}{}
 \footnotetext{$^\dag$ These authors contribute the paper equally.}
 \footnotetext{$^\ddag$ Corresponding Author$:$ astrophy0817@163.com}
 \institute{College of Physics and Engineering, Qufu Normal University, Qufu 273165, China\\
        \and
             Department of Astronomy and Atmospheric Sciences, Kyungpook National University, 1370 Sankyuk-dong, Buk-gu,
Daegu 702-701, Republic of Korea\\
        \and
            Shandong Provincial Key Laboratory of Laser Polarization and Information Technology, Qufu Normal University, Qufu 273165, China\\
\vs\no
   {\small Received~~20xx month day; accepted~~20xx~~month day}}

\abstract {Thanks to more and more gamma-ray bursts with measured redshift and extended emission detected by the recent space telescopes, it is urgent and possible to check whether those previous energy correlations still satisfy for the particular sample involving only the bursts accompanied by tail radiations. Using 20 long and 22 short bursts with extended emission, we find that the popular $\gamma$-ray energy correlations of the intrinsic peak energy versus the isotropic energy (Amati relation) and the intrinsic peak energy versus the peak luminosity (Yonetoku relation) do exist in either short or long bursts. However, these gamma-ray bursts with extended emissions are much better to be reclassified into two subgroups of E-I and E-II that make the above energy correlations more tight. As proposed by Zhang et al. (2018), the energy correlations can be utilized to distinguish these kinds of gamma-ray bursts in the plane of bolometric fluence versus peak energy as well. Interestingly, the peculiar short GRB 170817A belongs to the E-I group in the fluence versus peak energy plane, but it is an outlier of both Amati and Yonetoku relations even though the off-axis effect has been corrected. Furthermore, we compare the radiation features between the extended emissions and the prompt gamma-rays in order to search for their possible connections. Taking into account all these factors, we conclude that gamma-ray bursts with extended emission are still required to model with dichotomic groups, namely E-I and E-II classes, respectively, which hints that they might be of different origins.
}

   \authorrunning{X. L. Zhang et al. }            
   \titlerunning{Study of gamma-ray bursts with extended emission}  
   \maketitle

%
\section{Introduction}           
 \label{sec:intro}
The fascinating gamma-ray bursts(GRBs) are the fastest and most dynamic astronomical events in the universe \citep{Klebesadel+1973}.
The GRB durations ($T_{90}$) ranging from milliseconds to tens of minutes \citep{Zhang+2014} are usually used to express the lasting time of prompt $\gamma-rays$ \citep{Norris+1995}. According to the $T_{90}$ , GRBs can be traditionally classified into two types, namely long GRBs (LGRBs) with $T_{90} > 2s$ and short ones with $T_{90} < 2s$ \citep{Kouveliotou+1993} in the observer frame, and the bimodal distribution also exists in the rest frame \citep{Zhang+2008}.
This classification criterion had been confirmed by a number of observations \citep{Ghirland+2004,Paciesas+1999,Zhang+2016,Zitouni+2015,Zitouni+2018,Tarnopolski+19a,Tarnopolski+19b} while some other authors insisted that the number of subgroups in GRBs should be three \citep{Chattopadhyay+2007,Horvath+2016} or five \citep{Toth+2019,Chattopadhyay+2019}. However, a special kind of GRBs with extended emission (EE) component had been reported subsequently in many papers or catalogs \citep{Mazets+2004,Norris+2006,Kaneko+2015,Svinkin+2008}, which was found to confuse the classification scheme of long and short GRBs (SGRBs) according to the $T_{90}$ only \citep{Zhang+2016}. The extended emission had been thought to be produced by a relativistic wind extracting the rotational energy from a protomagnetar on a time-scale 10-100 s \citep{Metzger+2008}, the magnetar spin-down \citep{Zhang+2002,Fan+2006,Bucciantini+2012}, the process of fall-back accretion onto a newborn magnetar \citep{Gompertz+2014,Gibson+2017}, or a delayed energy injection causing the continued brightening of the early X-ray emissions as shown in GW170817/GRB 170817A \citep{Li+2018}.

Many authors argued that long GRBs are formed from the collapse of massive stars associated with
hypernovae \citep[e.g.,][]{Kinugawa+2019,Galama+1998,Hjorth+2003,Melandri+2014,Fruchter+2006}. Short GRBs are produced by the merger of either two neutron stars or a neutron star with a black hole \citep{Gompertz+2020,Li+1998,Fryer+1999,Popham+1998,Bulik+1998,Troja+2008,Wiggins+2018}. There are a number of empirical energy correlations for long bursts such as the
$\tau-L_p $ relation \citep{Norris+2000}, the $V-L_p $ relation \citep{Fan+20061,Reichart+2001}, the $N_{peak}-L_p $ relation \citep{Schaefer+2003}, the $\tau_{rel}-L_p$ relation \citep{Zhang+2006,Zhang+08}, the $\tau_{RT}-L_p $ relation \citep{Schaefer+2007}, the $E_{p,i}-\tau_{RT}-L_p $ relation \citep{Qi+2012}, the $L-T-E$
relation \citep{Xu+2012}, and the Liang-Zhang relation \citep{Liang+2005} etc, in which the intrinsic peak energy $E_{p,i}=(1+z)E_{p,o}$ versus the
isotropic energy $E_{iso}$ (hereafter Amati relation, \citealt{Amati+2002}) and the $E_{p,i}$ versus the peak luminosity $L_{p}$ (hereafter Yonetoku relation, \citealt{Yonetoku+2004}) are two frequently-studied ones. With the increasing number of short GRBs with known redshift, people found that at least parts of these above energy relations also hold for short GRBs. For example, \cite{Zhang+2018} (hereafter paper I) analyzed Swift/BAT and Fermi/GBM GRB data and found that the power law indexes of both Amati and Yonetoku relations of short GRBs are correspondingly consistent with those of long ones. This is however different from some early conclusions drawn by the limited data points of short GRBs \citep[e.g.,][]{Amati+2006, Amati+2012}. Despite decades of studying these sorts of energy relations, the underlying emission mechanisms still keep controversial \citep{Dainotti+2018,Ahlgren+2019}.

On the other hand, whether these kinds of energy relations also exist for the special EE bursts is an open question. In practice, the EE components following main peaks of a small fraction of GRBs have been identified not only in short bursts \citep{Ioka+05,Barthelmy+05,Norris+2006,Li+2020a,Li+2020b} but also in long ones \citep{Connaughton+02,Bostanci+13}. Moreover, \cite{Yu+2020} found that short GRBs with and without extended emissions are diversely distributed in the plot of peak flux versus fluence, which may indicate they are triggered by different binary coalescence mechanisms. Similarly, some long bursts also have softer gamma-ray emissions with very long timescale. In the recent years, more and more GRBs with softer EE tails are detected by the Swift satellite owing to its lower energy ranges. Therefore, the primary task of the paper is to test the existent possibilities and the consistency of the Amati and Yonetoku relations of the EE bursts with those obtained with normal GRBs previously. Additionally, we shall check how to reclassify these EE bursts would be more appropriate according to their diverse energy correlations. It is noticeable that GRB 170817A as the first gravitational-wave-associated short GRB with EE will be paid more attentions on its classification. Sample selection and data reduction methods are displayed in Sec. \ref{sec:DM}. Our results are presented in Sec. \ref{sec:result}. We will end with conclusions in Sec. \ref{sec:Conclusion}.


\section{Data and Methods } \label{sec:DM}

Firstly, we collect the GRBs with EE and redshift reported in literatures between July 2005 and August 2017
\citep{Norris+2006,Gompertz+2013,van+2014,Kaneko+2015,Zhang+2016,Gibson+2017,Kisaka+2017,Yu+2020}. In order to reduce the sampling selection effect, we chose not only short GRBs but also long bursts to constitute our complete GRB sample including EE bursts only in this paper. \cite{Kisaka+2017} proposed a phenomenological formula consisting of two functions to identify the EE components and got 65 GRBs with EE, of which less than half have the measured redshifts. However, parts of them could not show the obvious EE segments in view of their light curves of multi-energy bands, especially in lower energy channels. To ensure the sampling reliability, we have double-checked the light curves with a criterion of signal-to-noise ($S/N$) larger than 2 to judge the EE segments for the EE candidates taken from literatures. In total, 42 EE GRBs with known redshift are chosen to compose our sample. Of the 42 EE bursts, 20 long and 22 short bursts are included, 28 and 14 GRBs are respectively detected by Swift/BAT and other satellites. It happens that the EE GRB sample also consists of 20 E-I and 22 E-II bursts. Note that the E-I and E-II GRBs are not equal to the short and long ones, correspondingly. (see the definition in Sec. \ref{sec:result--2} for details). The physical parameters are listed in Table \ref{tab1}, where Column 1 gives the GRB name, Column 2 lists the duration $ T_{90}$, Column 3 gives the cosmological redshift, Columns 4-6 respectively represent the observed peak energy $E_{p,o}$, and two spectral indexes ($\alpha$ and $\beta$) of the GRB $\nu F_\nu$ spectrum, Columns 7 and 8 provide the observed energy fluence  $S_\gamma$ in units of erg/cm$^2$ and peak photon flux $P_{\gamma}$ in units of ph/cm$^2$/s, Column 9 and 10 show the energy bands from $E_{min}$ to $E_{max}$ of detectors and their corresponding $K$-correction factors $K_c$ from the observer frame to the source frame in energy band 1-10000 keV, individually. The relevant references are inserted in Column 11. Finally, E-I and E-II in Column 12 denote the detailed types of the EE GRBs based on the different energy correlations they are matching.

Subsequently, we will use the selected sample of EE bursts to study their potential energy correlations that can be applied to classify them into different subgroups. The methods and steps are completely same as in our previous paper I \citep{Zhang+2018}. In addition, we comparatively investigate the radiation properties of the EE components and the main peak emissions of the prompt $\gamma$-rays for distinct classes of GRBs with EE. Hopefully, we shall find some possible connections of the EE segments with their corresponding main peaks in order to explore the EE origins. For this purpose, the times and photon fluxes when the EE parts ($t_{p,EE}$ and $F_{p,EE}$) and the main bursts ($t_{p,main}$ and $F_{p,main}$) peak separately are measured and compared. Note that two peak times are recorded from the trigger time of a detector and the peak fluxes are measured for the mask-weighted light curves. Especially, two variables, $t_{p,EE}$ and $F_{p,EE}$, have been estimated from the lower energy channel where the EE components are usually identified and relatively softer than the main bursts. To ensure the EE segments to be reliably measured, the selection criterion of $S/N\geq3$ has been adopted. In this way, we pick out 10 short and 19 long GRBs to study the relationships of timescales, intensities together with energy correlations of the EE portions. We need to point out that 10 E-I and 19 E-II bursts are also involved for this comparative study. It is however a coincidence that the numbers of different kinds of bursts are unexpectedly equal.

\begin{table}
\bc
\begin{minipage}[]{100mm}
\caption[]{Physical parameters of GRBs with EE\label{tab1}}\end{minipage}
\setlength{\tabcolsep}{5.5pt}
\small
 \begin{tabular}{lccccccccccl}
  \hline\noalign{\smallskip}
GRB&$T_{90}$ & $z$ & $E_\text{p}$ & $\alpha$ & $\beta$ & $S_\gamma$ & $P_{\gamma}$ & $E_{min}-E_{max}$ & $K_c$  & $Ref$& Type\\
 & $(s)$ & & (keV) & & & (erg/cm$^2)$ &(ph/cm$^2$/s) & (keV) &&&  \\
(1)&(2)&(3)&(4)&(5)&(6)&(7)&(8)&(9)&(10)&(11)&(12)\\
  \hline\noalign{\smallskip}
050724$^\star$ &96&0.257&78.91$\pm$8.0&-2.02&-&8.90E-07&3.35&15-150&5.37 &[1,12]&E-II \\
051016B$^\star$ &4&0.9364&20.42$\pm$5.34&-1.588&-&1.67E-07&0.685&15-350&2.26 &[1,13]&E-II \\
051221A&1.4&0.547&402$\pm$93&-1.08&-&3.20E-06&12.1&20-2000&1.07 &[4,10]&E-I \\
051227$^\star$&114.6&0.8&332.01$\pm$211.02&-1.41&-&7.09E-07&0.95&15-350&1.77 &[1,3]&E-I \\
060306$^\star$ &60.94&1.559&69.38$\pm$13.67&-1.254&-&2.45E-06&6.41&15-350&1.35 &[1,13]&E-II \\
060313&0.74&1.7&837.41$\pm$438.12&-0.61&-&1.14E-06&10.85&15-350&3.68 &[1,3]&E-I \\
060614$^\star$&108.7&0.125&393.02$\pm$250.96&-2.23&-&1.88E-05&11.39&15-350&7.84 &[1,3]&E-II \\
060801$^\star$&0.49&1.13&620.22$\pm$342.95&0.28&-&7.84E-08&0.75&15-350&3.47 &[1,3]&E-I \\
060814$^\star$ &145.3&0.84&302.37$\pm$127.18&-1.412&-&2.39E-05&8.38&15-350&1.72 &[1,13]&E-II \\
061006$^\star$&129.9&0.4377&664$\pm$227&-0.62&-&3.57E-06&5.3&20-10000&1.01 &[5,12]&E-I \\
061201&0.76&0.111&873$\pm$458&-0.36&-&5.32E-06&3.55&20-3000&1.02 &[6,12]&E-I \\
061210$^\star$&85.3&0.41&544.04$\pm$309.56&-1.56&-&1.10E-06&2.78&15-350&2.20 &[1,12]&E-I \\
070223$^\star$ &100&1.6295&54.44$\pm$14.45&-1.48&-&1.98E-06&0.491&15-350&1.57 &[1,13]&E-II \\
070506$^\star$ &4.3&2.31&55.12$\pm$11.29&-0.768&-&2.22E-07&0.948&15-350&1.24 &[1,13]&E-II \\
070714B$^\star$&64&0.92&164.87$\pm$73.13&-1.15&-&7.23E-07&2.75&15-350&1.31 &[1,3]&E-II \\
070724A&0.4&0.457&82$\pm$5&-1.15&-&3.00E-08&0.94&15-150&1.56 &[7,12]&E-I \\
071227$^\star$&1.8&0.383&1000$\pm$100&-0.7&-&1.60E-06&1.68&20-1000&1.64 &[8,12]&E-I \\
080123$^\star$&115&0.495&44.93$\pm$4.49&-1.99&-&5.52E-07&1.43&15-350&2.63 &[1,12]&E-II \\
080603B$^\star$ &60&2.69&74.94$\pm$10.86&-1.21&-&2.98E-06&4.72&15-350&1.32 &[1,13]&E-II \\
080905A$^\star$&1&0.128&311.2$\pm$100&0.12&-2.35&8.51E-07&6.32&10-1000&1.51 &[1,12]&E-I \\
080905B$^\star$ &128&2.374&256.10$\pm$65.06&-1.579&-2.29&2.75E-06&1.03&15-350&1.80 &[1,13]&E-II \\
090426$^\star$&1.2&2.609&55.09$\pm$27&-1.11&-&1.76E-07&2&15-150&1.49 &[1,9]&E-II \\
090510$^\star$&0.3&0.903&4302$\pm$483.2&-0.86&-2.58&3.37E-06&40.95&10-1000&3.94 &[2,11]&E-I \\
090530$^\star$ &40.46&1.266&92.14$\pm$30.56&-1.078&-&1.33E-06&3.68&15-350&1.23 &[1,13]&E-II\\
090927$^\star$ &2.2&1.37&61.95$\pm$19.12&-1.301&-&2.97E-07&1.85&15-350&1.4 &[1,13]&E-II \\
100117A&0.3&0.915&327.22$\pm$52.91&-0.1&-6.3&9.26E-08&0.96&10-1000&1.02 &[1,2]&E-I \\
100625A&0.33&0.452&482.13$\pm$61.93&-0.59&-12.24&2.32E-07&2.54&10-1000&1.12 &[1,2]&E-I \\
100704A$^\star$ &197.5&3.6&381.75$\pm$80.77&-1.655&-2&8.91E-06&5.1&15-350&2.05 &[1,13]&E-II \\
100724A&1.4&1.288&42.5$\pm$15.18&-0.51&-&1.41E-07&1.56&15-150&1.29 &[1,12]&E-II\\
100814A$^\star$ &174.5&1.44&312.96$\pm$188.9&-1.331&-2.44&1.47E-05&3.05&15-350&1.7 &[1,13]&E-II \\
100906A$^\star$ &114.4&1.727&138.37$\pm$36.45&-1.722&-1.86&1.89E-05&11.1&15-350&1.84 &[1,13]& E-II\\
101219A&0.6&0.718&490$\pm$103&-0.22&-&3.60E-06&4.2&20-10000&1.01 &[9,10]&E-I \\
111117A&0.47&2.211&370$\pm$37&-0.69&-&6.70E-07&2.8&15-150&3.84 &[1,10]&E-I \\
120804A$^\star$&0.81&1.3&116.18$\pm$39.82&-0.97&-&8.66E-07&10.64&15-350&1.19 &[1,3]&E-II \\
131004A$^\star$&1.54&0.71&118.1$\pm$29.7&-1.36&-22.09&5.09E-07&9.82&10-1000&1.17 &[1,2]&E-II \\
150120A&1.2&0.46&130$\pm$50&-1.43&-1.65&4.17E-07&4.94&10-1000&2.31 &[1,2]&E-II\\
150423A&0.22&1.39&120$\pm$35&0.43&-&6.30E-08&2.6&15-150&1.42 &[1,12]&E-I\\
150424A&91&0.3&47.06$\pm$6.64&-0.49&-2.19&1.50E-06&12&10-1000&1.12 &[1,2]&E-II\\
160410A$^\star$&8.2&1.717&495.3$\pm$232.9&-1.11&-&1.15E-06&0.34&15-350&2.07 &[1,3]&E-I\\
160624A&0.2&0.483&1168$\pm$546.5&-0.63&-3.65&1.21E-07&6.39&10-1000&1.84 &[1,2]&E-I\\
160821B$^\star$&0.48&0.16&46.32$\pm$5.38&-0.12&-&1.03E-07&1.68&15-150&1.19 &[1,2]&E-I\\
170817A$^\star$ &2.05&0.009783&214.7$\pm$56.6&-0.60&-&2.79E-07&3.73&10-1000&1.00 &[2,12]&E-I\\
  \noalign{\smallskip}\hline
\end{tabular}
\ec
\tablecomments{1\textwidth}{[1] https://swift.gsfc.nasa.gov/results/batgrbcat/; [2] https://heasarc.gsfc.nasa.gov/W3Browse/fermi/fermigtrig.html;
[3] Butler et al. 2007; [4] Golenetskii et al. 2005; [5] Golenetskii et al. 2006a; [6] Golenetskii et al. 2006b; [7] Golenetskii et al. 2006c;
[8] Golenetskii et al. 2007; [9] Goldstein et al. 2010£»[10] Fong et al. 2015; [11] Razzaque 2010; [12] Goldstein et al. 2017; Zhang et al.(2018); [13] Zhang et al.(2016). Symbol stars denote those GRBs with brighter EE components at a level of $S/N\geq3$.}
\end{table}


\section{Result}\label{sec:result}

\subsection{Parameter distributions}\label{sec:result--1}
\begin{figure}
\begin{minipage}[t]{1\linewidth}
\begin{center}
\includegraphics[width=0.8\textwidth]{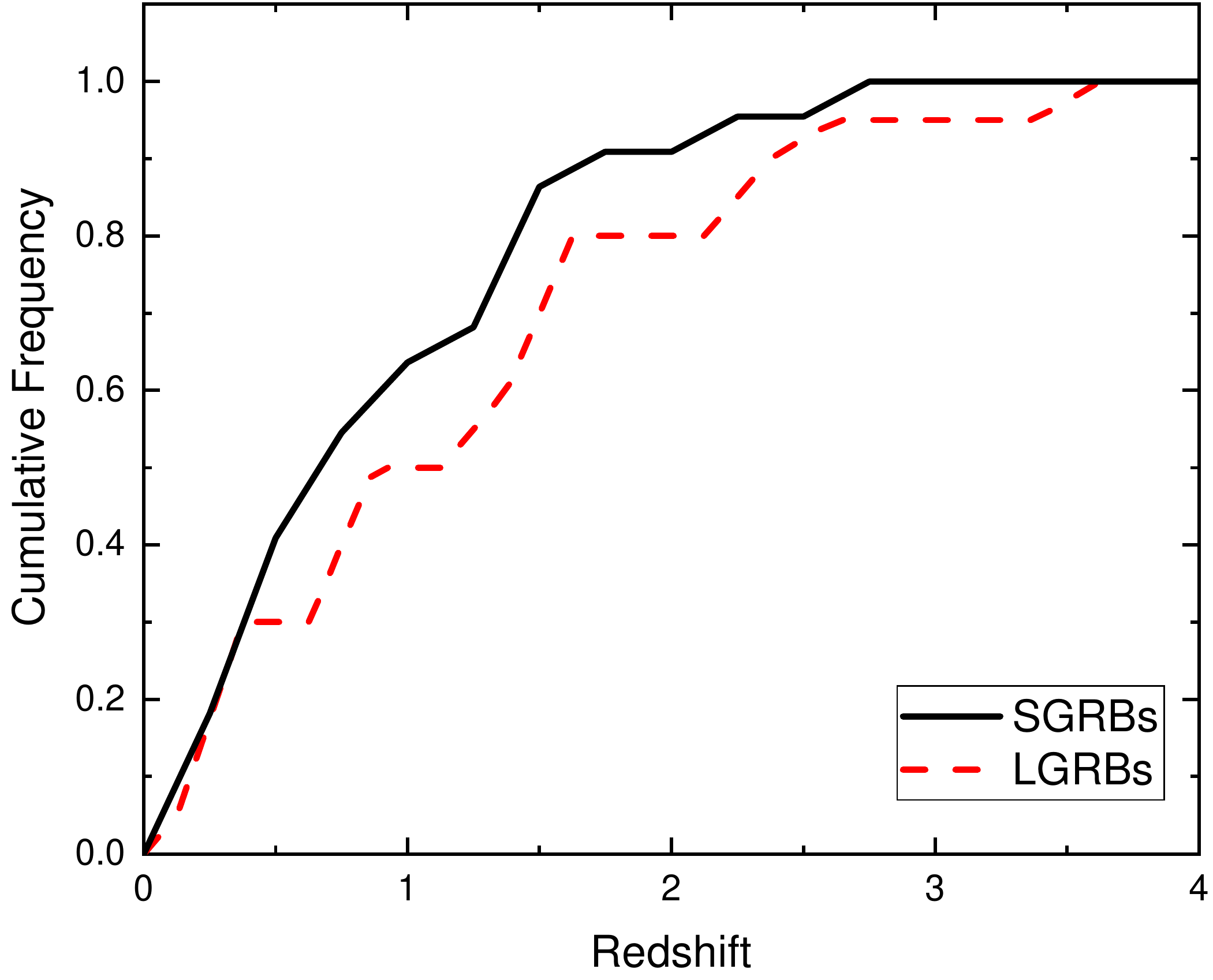}
\end{center}
\end{minipage}
\begin{minipage}[t]{1\linewidth}
\begin{center}
\includegraphics[width=0.8\textwidth]{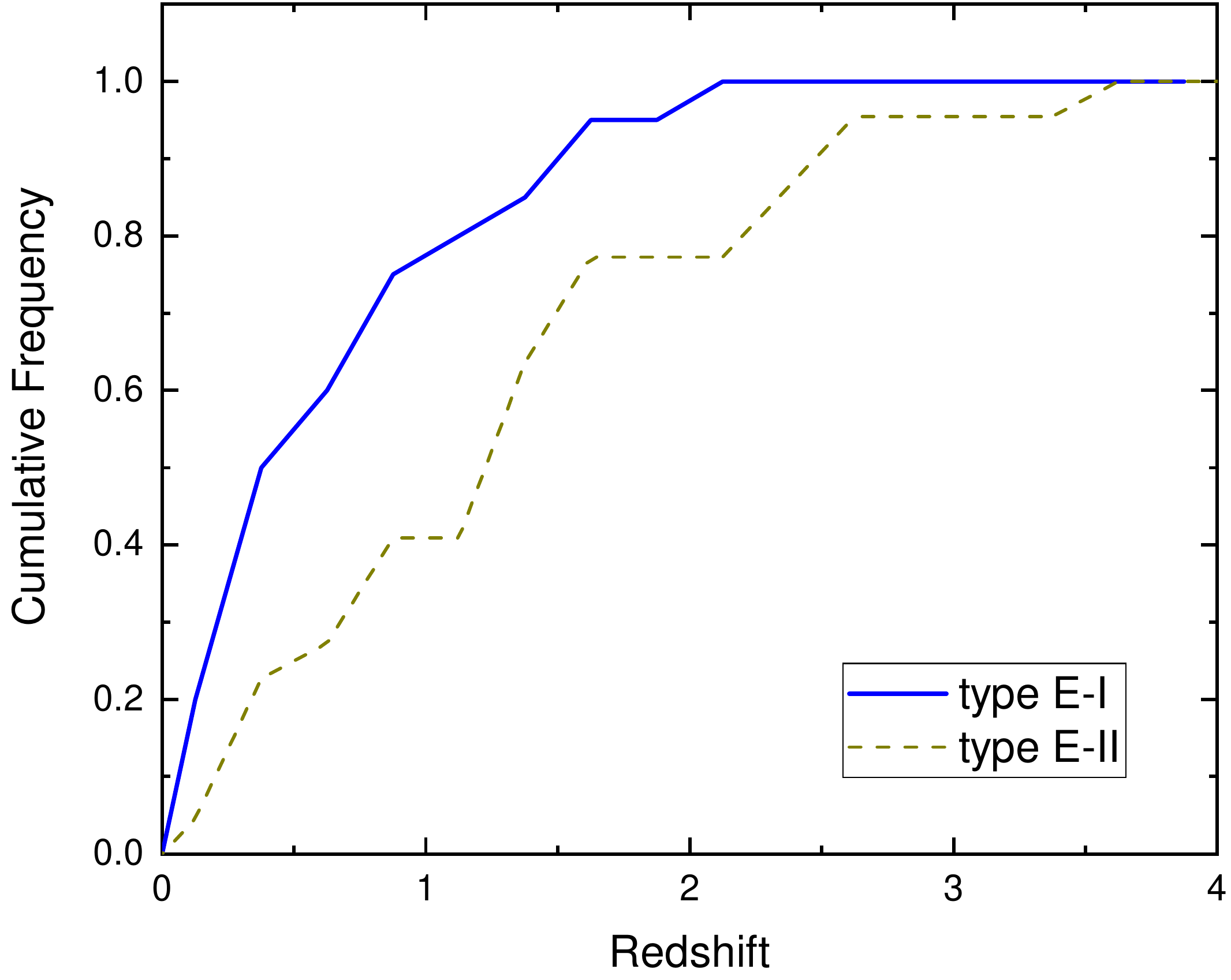}
\caption{Top panel: Cumulative probability distributions of redshifts for short (solid line) and long (dashed line) EE GRBs. Bottom panel: Cumulative probability distributions of redshifts for E-I (solid line) and E-II (dashed line) GRBs with EE.\\
   (A color version of this figure is available in the online journal)}
\label{sec:fig1}
\end{center}
\end{minipage}

\end{figure}

The redshift distributions of different EE bursts in our sample are shown in Figure \ref{sec:fig1}, where the median redshifts are $z=$0.71, 1.1, 0.52 and 1.29 for short, long, E-I and E-II GRBs, respectively. It is noticeable that the redshift differences between E-I and E-II GRBs are comparably larger than those between short and long bursts on a whole. A K-S test to the redshift distributions of short and long bursts returns the statistic $D=0.31$ less than the critical value of $D_{\alpha'}(n_{1},n_{2})=0.42$ and the \textit{p}-value of 0.2 at a significance level $\alpha'=0.05$ for $n_1=20$ and $n_2=22$, showing short and long GRBs share with the same redshift distribution. While a K-S test to the redshift distributions of E-I and E-II bursts returns the statistic $D=0.43>D_{\alpha'}(n_{1},n_{2})=0.42$ and the \textit{p}-value of 0.03 for $\alpha'=0.05$, which indicates that the redshifts of E-I and E-II bursts are drawn from different parent distributions.


\begin{figure}
\begin{minipage}[t]{1\linewidth}
\begin{center}
\includegraphics[height=10cm,width=12.0cm]{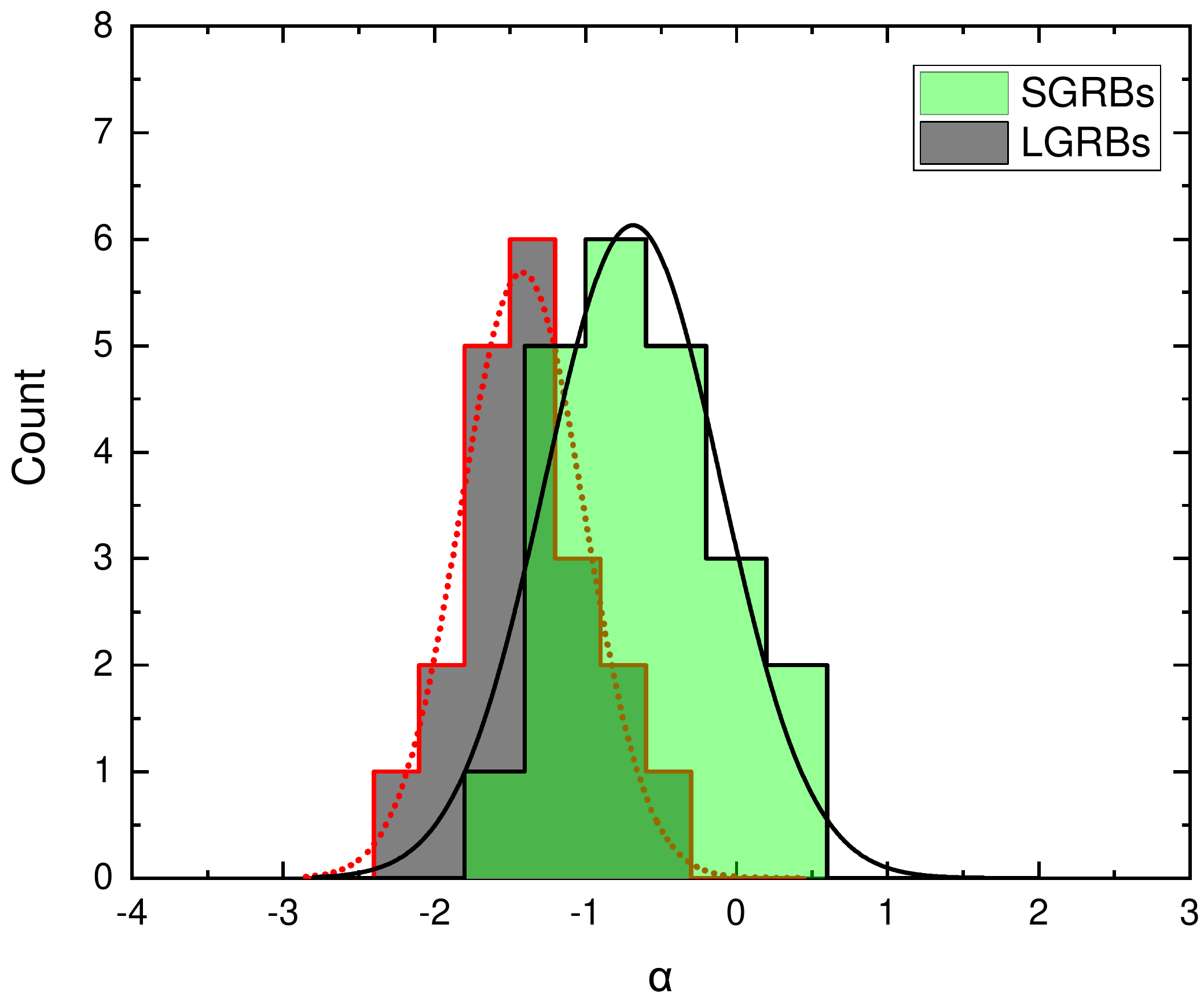}
\end{center}
\end{minipage}
\begin{minipage}[t]{1\linewidth}
\begin{center}
\includegraphics[height=10cm,width=12.0cm]{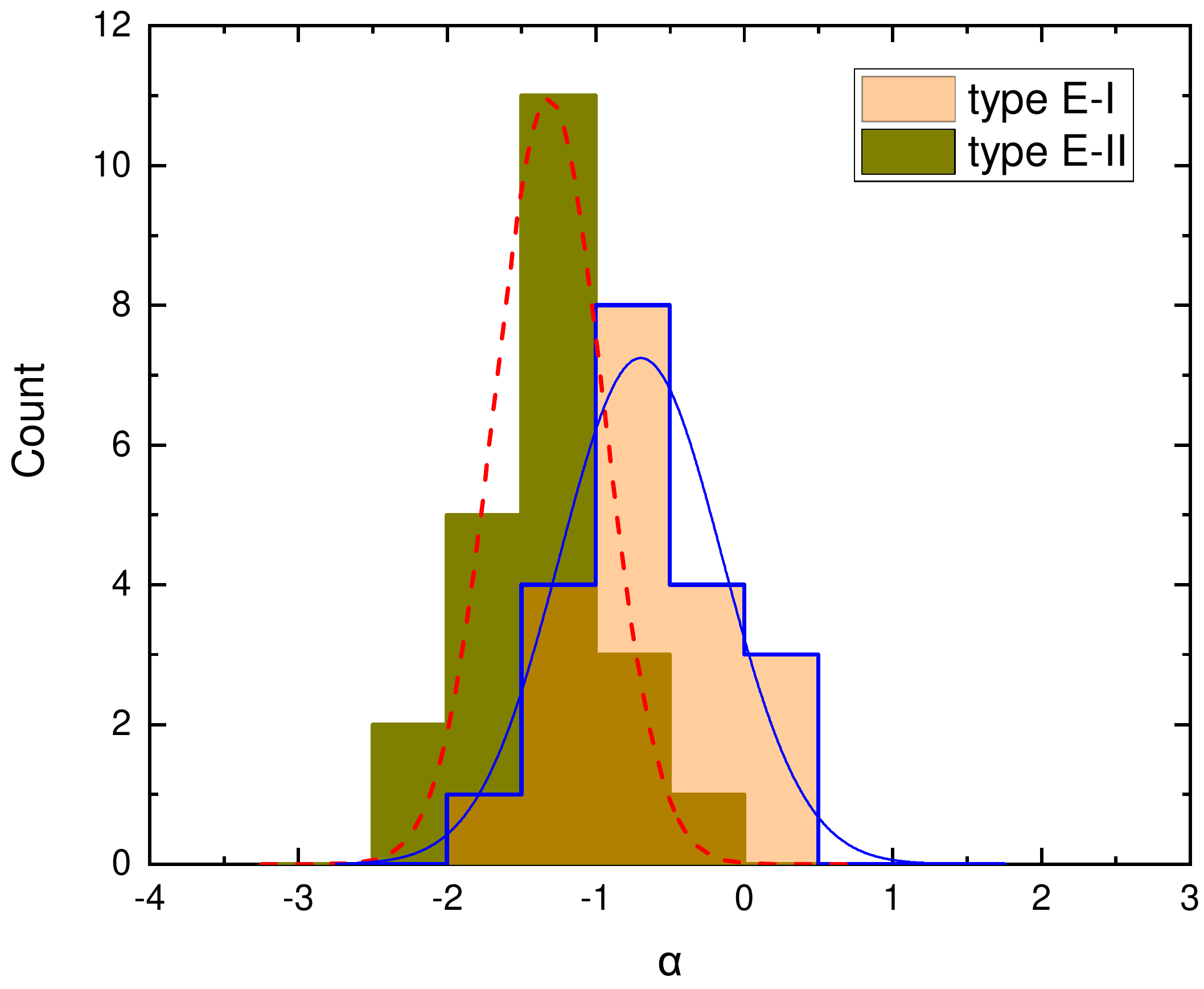}
\caption{Top panel: Distributions of low-energy spectral index of $\alpha$ in Band function for short (green) and long (gray) GRBs. Bottom panel: Distributions of low-energy spectral index of $\alpha$ in Band function for E-I (light orange) and E-II (dark yellow) GRBs. The different lines are the best fits to the histograms with a Gaussian function.\\
 (A color version of this figure is available in the online journal)}
\label{sec:fig2}
\end{center}
\end{minipage}
\end{figure}

\begin{figure}
\begin{minipage}[t]{1\linewidth}
\begin{center}
\includegraphics[height=10cm,width=12.0cm]{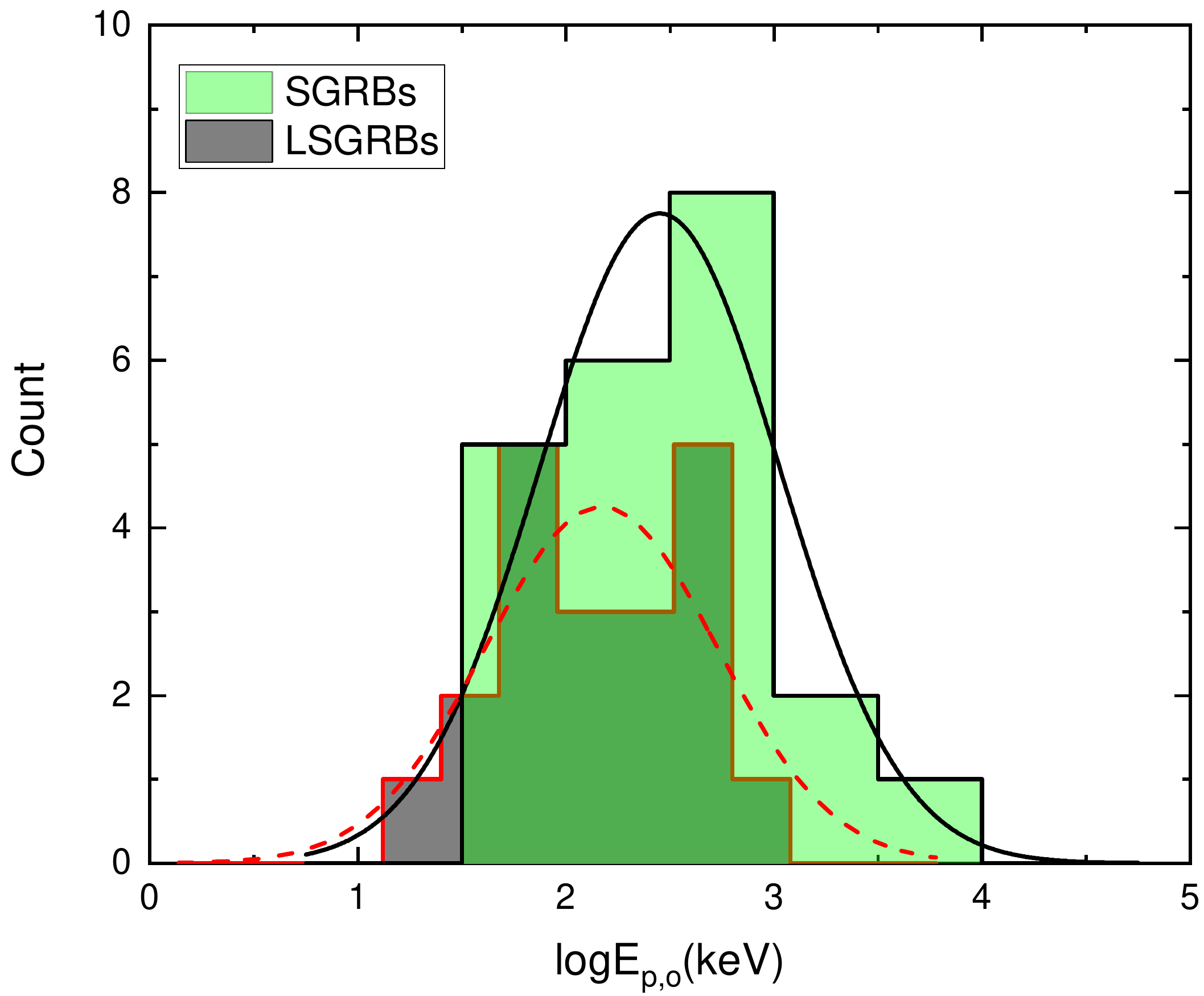}
\end{center}
\end{minipage}
\begin{minipage}[t]{1\linewidth}
\begin{center}
\includegraphics[height=10cm,width=12.0cm]{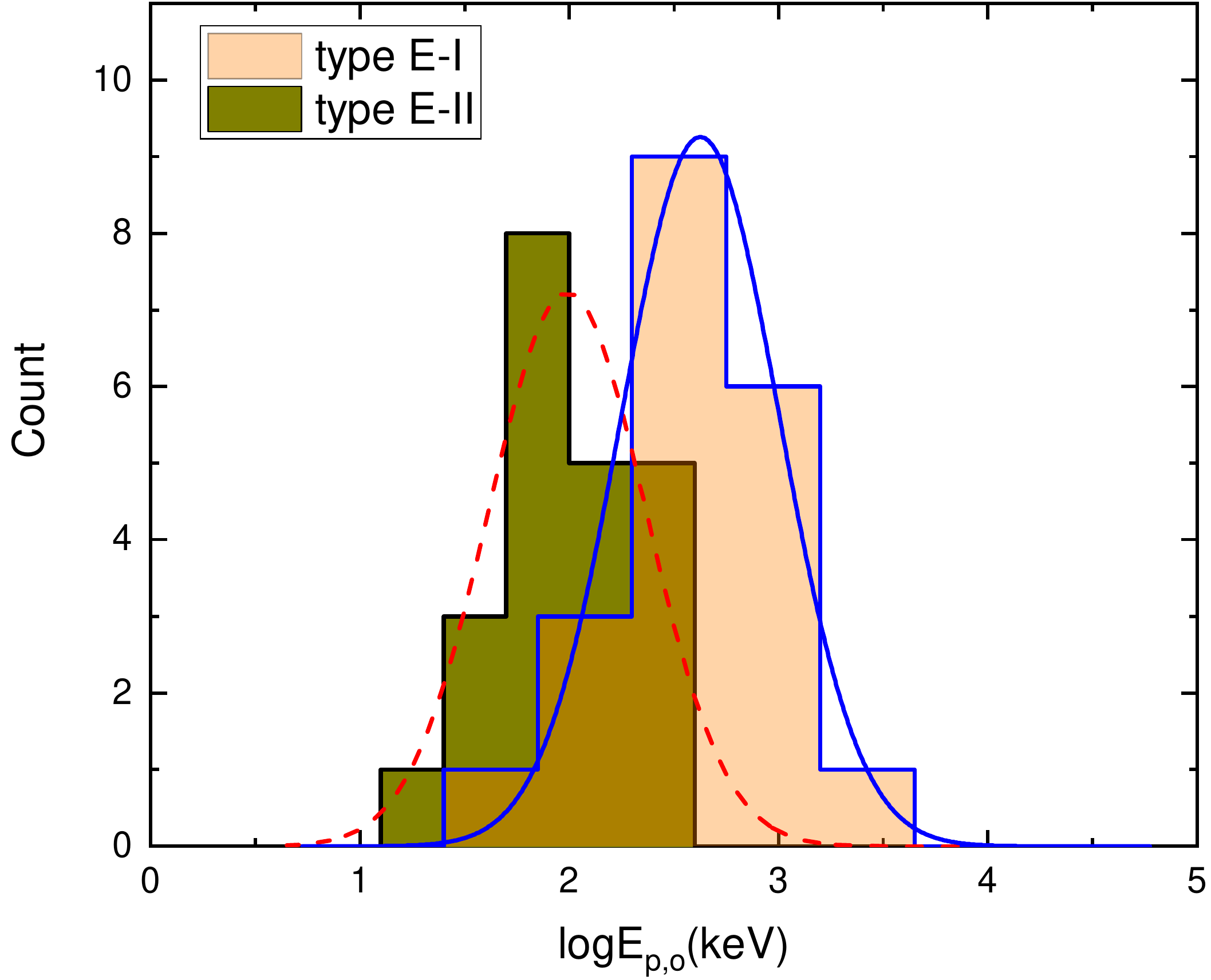}
\caption{Top panel: Distributions of $E_{p,o}$ in Band function for short (green) and long (gray) GRBs. Bottom panel: Distributions of $E_{p,o}$ in Band function for E-I (light orange) and E-II (dark yellow) GRBs. The distinct lines are the best fits to the histograms with a lognormal function.\\
   (A color version of this figure is available in the online journal)}
\label{sec:fig3}
\end{center}
\end{minipage}
\end{figure}

\begin{figure}
\begin{minipage}[t]{1\linewidth}
\begin{center}
\includegraphics[height=10cm,width=12.0cm]{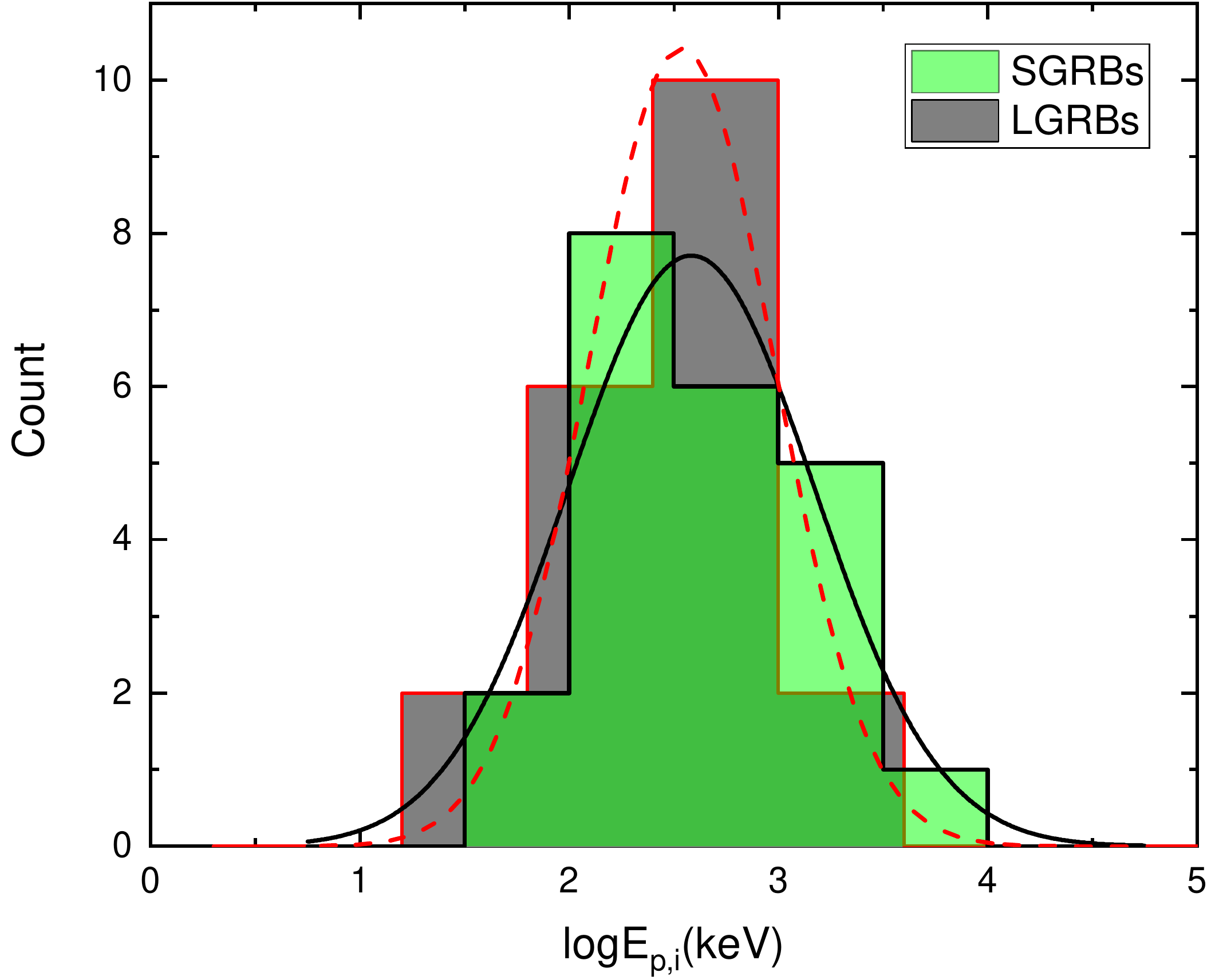}
\end{center}
\end{minipage}
\begin{minipage}[t]{1\linewidth}
\begin{center}
\includegraphics[height=10cm,width=12.0cm]{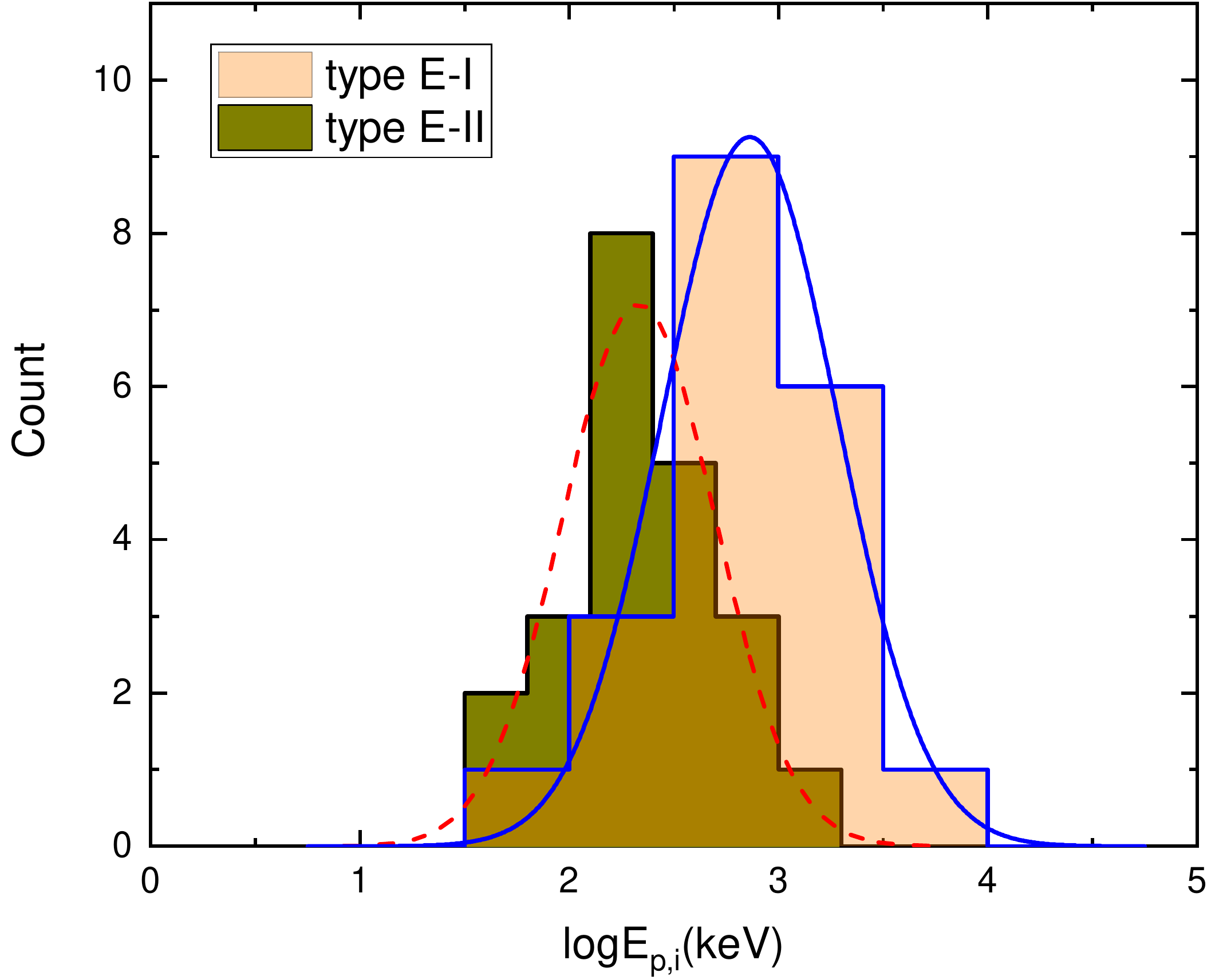}
\caption{Top panel: Distributions of the intrinsic peak energy of $E_{p,i}$ for short (green) and long (gray) GRBs. Bottom panel: Distributions of the intrinsic peak energy of $E_{p,i}$ for E-I (light orange) and E-II (dark yellow) GRBs. The distinct lines are the best fits to the histograms with a lognormal function.\\
   (A color version of this figure is available in the online journal)}
\label{sec:fig4}
\end{center}
\end{minipage}
\end{figure}

Figure \ref{sec:fig2} displays the distributions of the low-energy spectral index of $\alpha$ with mean values of -0.69, -1.41, -0.70, -1.31 and scatters of 0.58, 0.41, 0.55, 0.37 for the short, long, E-I, E-II GRBs, respectively. The K-S tests give $D=0.58$ and $P=8.5\times10^{-4}$ between short and long GRBs, and $D=0.61$ and $P=3.7\times10^{-4}$ between E-I and E-II GRBs, which demonstrates that they all are differently distributed. In Figure \ref{sec:fig3} we compare the $E_{p,o}$ distributions and get $D_{\alpha'}(n_{1},n_{2})=0.26$ with $P=0.41$ between short and long bursts and $D_{\alpha'}(n_{1},n_{2})=0.66$ with $P=7.8\times10^{-5}$ between E-I and E-II bursts which shows that the observed peak energies of E-I and E-II GRBs significantly have diverse distributions. However, the $E_{p,o}$ distributions of short and long GRBs are  statistically same as some previous authors found for BATSE and Swift bursts \citep{Ghirlanda04,Zhang2020}. The mean $E_{p,o}$ values of short, long, E-I and E-II GRBs are respectively $281.8^{+55.9}_{-46.2}$, $147.9^{+38.3}_{-30.4}$, $422.7^{+18.2}_{-12.9}$ and $97.7^{+11.9}_{-10.6}$ keV. The $E_{p,i}$ distributions in Figure \ref{sec:fig4} are very similar to Figure \ref{sec:fig3} and also show that short and long bursts are taken from the same parent distribution while E-I and E-II GRBs are differently distributed. We notice that the average $E_{p,i}$ value of type E-I GRBs is still larger than that of type E-II GRBs in the rest frame. Nevertheless, the mean $E_{p,i}$ values are $380.2^{+66.5}_{-46.2}$ and $346.7^{+24.8}_{-23.1}$ for short and long GRBs respectively and a K-S test gives $D_{\alpha'}(n_{1},n_{2})=0.22$ with $P=0.61$ showing their $E_{p,i}$ distributions are uniform.

\subsection{Spectrum-energy relations}\label{sec:result--2}
\begin{figure}
\begin{minipage}[t]{1\linewidth}
\begin{center}
\includegraphics[height=10cm,width=12.0cm]{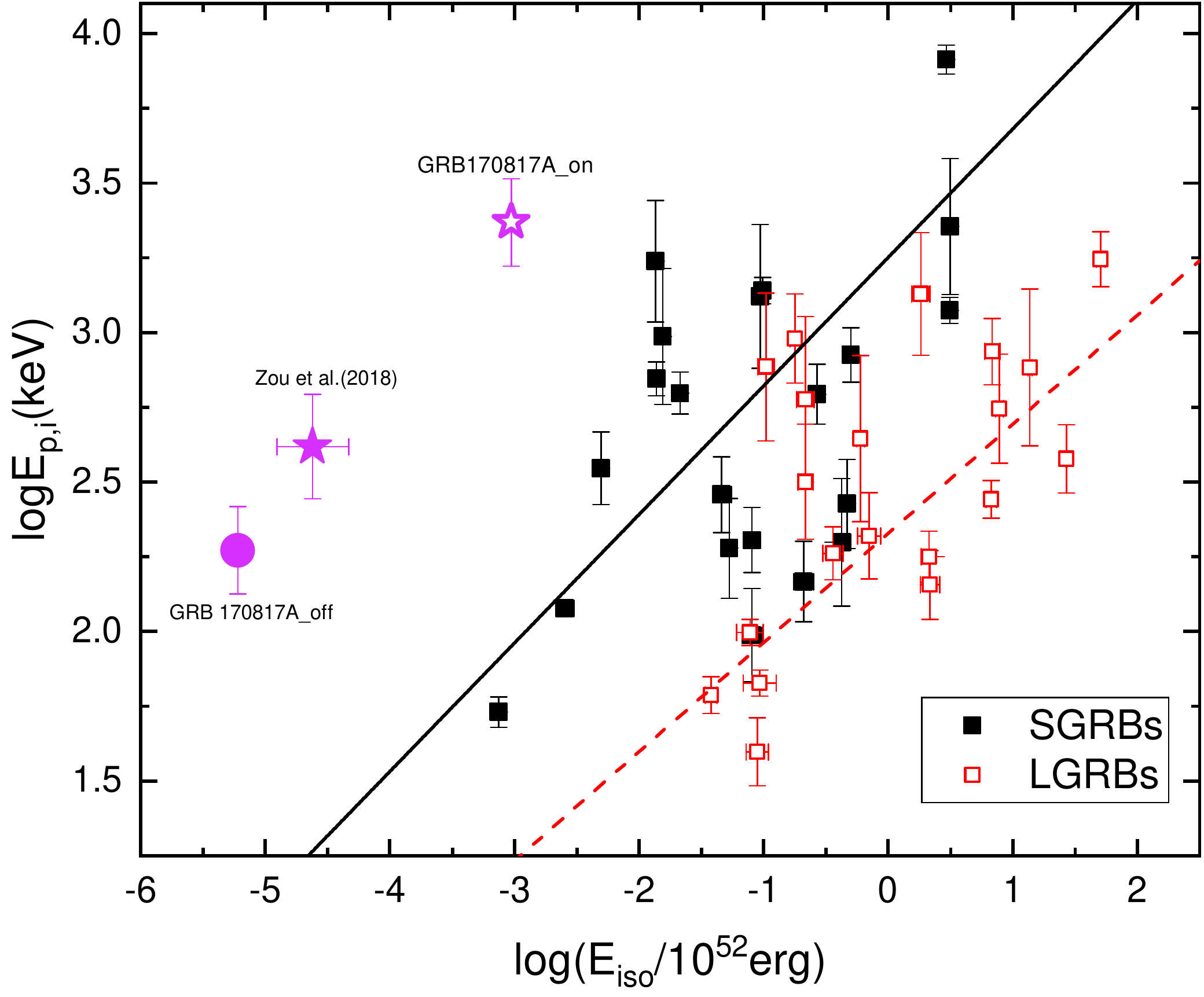}
\end{center}
\end{minipage}
\begin{minipage}[t]{1\linewidth}
\begin{center}
\includegraphics[height=10cm,width=12.0cm]{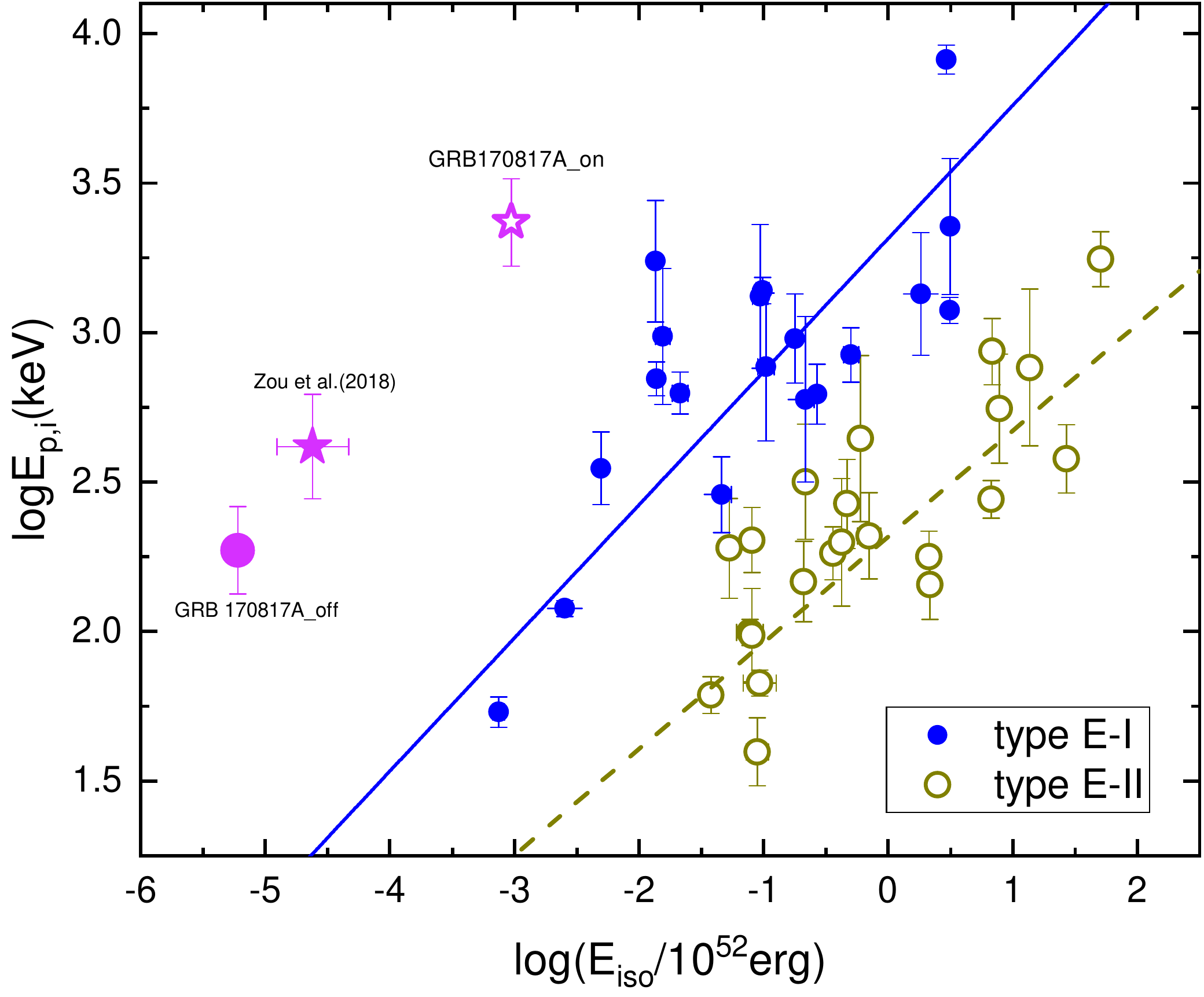}
\caption{The energy relations of $E_{p,i}$ versus $E_{iso}$ in logarithmic scale for short (filled square) and long (empty square) GRBs in upper panel and for E-I (filled circle) and E-II (empty circle) GRBs in lower panel. The straight lines stand for the best fits to data. For the peculiar SGRB 170817A, the filled large circle shows the off-axis measurements, while the on-axis parameters given by Zou et al. (2018) and this work are denoted by the filled and the empty stars, respectively (see text for details).\\
(A color version of this figure is available in the online journal)}
\label{sec:fig5}
\end{center}
\end{minipage}
\end{figure}

Following our paper I, we use the data in Table \ref{tab1} to calculate the isotropic energy $E_{iso}=4\pi D_{l}^{2} S_{bolo}(1+z)^{-1}$ and the peak luminosity $L_p=4\pi D_{l}^{2}P_{bolo}$, where $D_l$ is the cosmological distance, $S_{bolo}=K_c S_{\gamma}$ and $P_{bolo}=K_cP_{\gamma}$ are bolometric fluence and flux transferred from the observed fluence $S_{\gamma}$ and flux $P_{\gamma}$ with a \textit{K}-correction factor of $K_c$ \citep{Zhang+2018}. Figure \ref{sec:fig5} shows the Amati relations of $E_{p,i}\sim C_1 E_{iso}^{\eta_{1}}$ for the above four EE GRB groups in the rest frame. They can be individually written as
\begin{equation}
 E_{p,i}=1783.61^{+527.5}_{-407.3}\left(\frac{E_{iso}}{10^{52}erg}\right)^{0.43\pm0.06 } (keV) \label{equation:1}
\end{equation}
for 21 SGRBs and
\begin{equation}
 E_{p,i}=212.82^{+32.4}_{-28.2}\left(\frac{E_{iso}}{10^{52}erg}\right)^{0.37\pm0.06 } (keV)\label{equation:2}
\end{equation}
for 20 LGRBs. However, short and long GRBs are moderately overlapped and dispersedly distributed in the plane of $E_{p,i}$ vs. $E_{iso}$. If redividing these EE bursts into E-I and E-II subgroups as shown in the lower panel of Figure \ref{sec:fig5}, one can obtain two more tight Amati relations with smaller standard deviations to be
\begin{equation}
 E_{p,i}=2062.76^{+552.9}_{-436.0}\left(\frac{E_{iso}}{10^{52}erg}\right)^{0.45\pm0.05 } (keV)\label{equation:3}
\end{equation}
for 19 E-I GRBs and
\begin{equation}
 E_{p,i}=207.60^{+23.5}_{-21.1}\left(\frac{E_{iso}}{10^{52}erg}\right)^{0.36\pm0.04 } (keV)\label{equation:4}
\end{equation}
for 22 E-II GRBs, respectively, which conversely shows that the E-I/II classification could be more physical. It is noteworthy that GRB 170817A has not been utilized during the above fits. All the fitting parameters are listed in Table \ref{tab2} where one can find that the power-law indexes are marginally consistent with each other and the energy correlations of short and long bursts is much closer to those of E-I and II GRBs. In addition, the fitted $\eta_{1}$ values are surprisingly coincident with those obtained by paper I for 31 short and 252 long GRBs with lower $E_{p,o}$ mainly observed by Swift/BAT, but slightly smaller than the previous value of $\eta_{1} \simeq0.5$ \citep[e.g.][]{Amati+2002,Amati+2006,Amati+2019}. This hints that the Amati relation might evolve with the peak energy. In particular, we find that the peculiar GRB 170817A always violate the new-built Amati relations even though the off-axis effect has been corrected according to the method used by \cite{Zou+18}. To perform the off-axis corrections for GRB 170817A, we adopt the viewing angle of $\theta_v$= 0.53 radians, the half-opening jet angle $\theta_j=$0.1 radians from \cite{Hajela+19} and the initial Lorentz factor $\Gamma=8$ \citep{Salafia+18}. Its on-axis energies are $E_{p,i,on}=2713.4\pm715.3$ keV and $E_{iso,on}=(9.23\pm0.56)\times10^{48}$ erg that are correspondingly about one order of magnitude larger than the   those estimated by \cite{Zou+18}, where $\Gamma=13.4$ and $\theta_v=0.175$ radians had been assumed.

\begin{figure}
\begin{minipage}[t]{1\linewidth}
\begin{center}
\includegraphics[height=10cm,width=12.0cm]{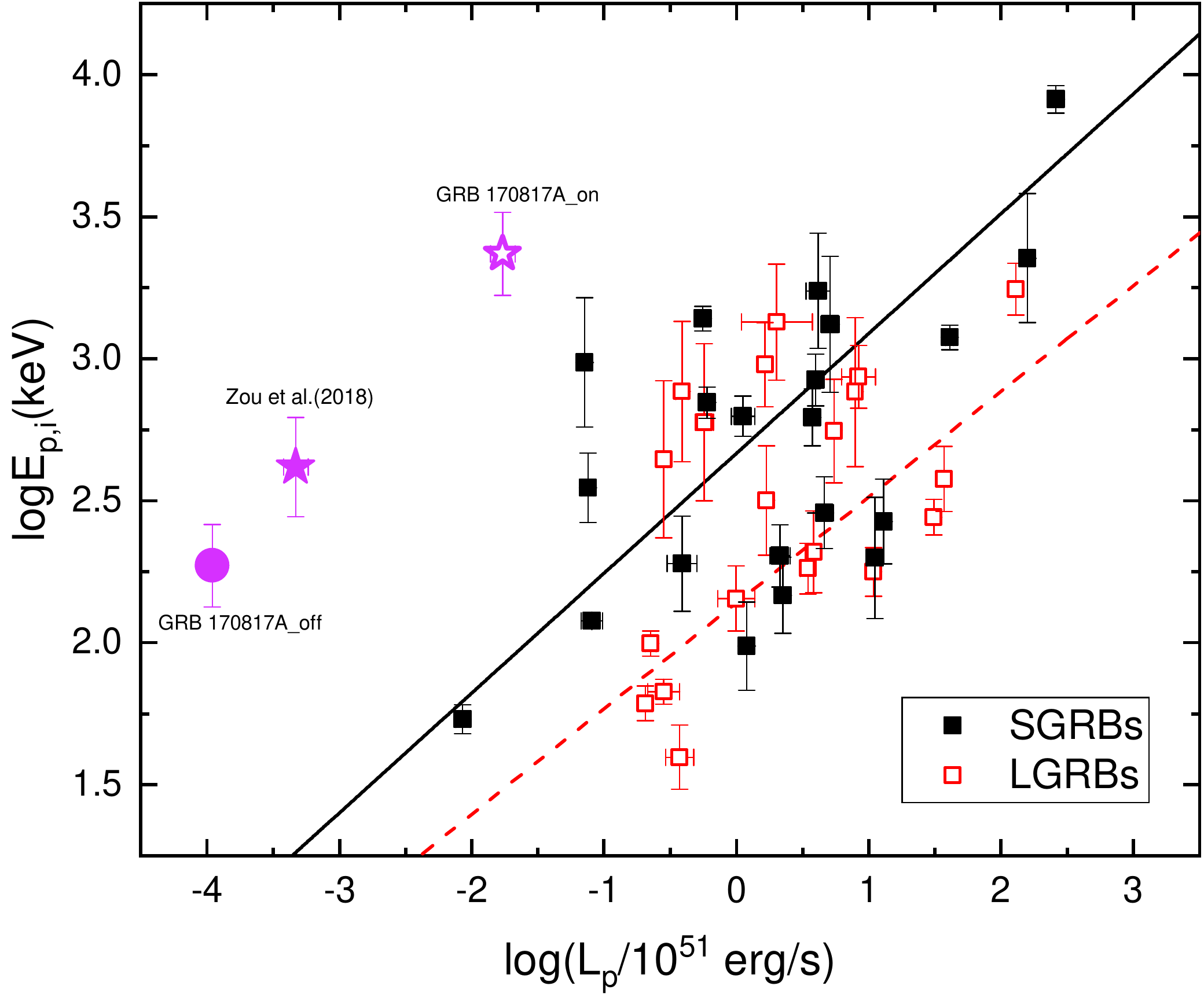}
\end{center}
\end{minipage}
\begin{minipage}[t]{1\linewidth}
\begin{center}
\includegraphics[height=10cm,width=12.0cm]{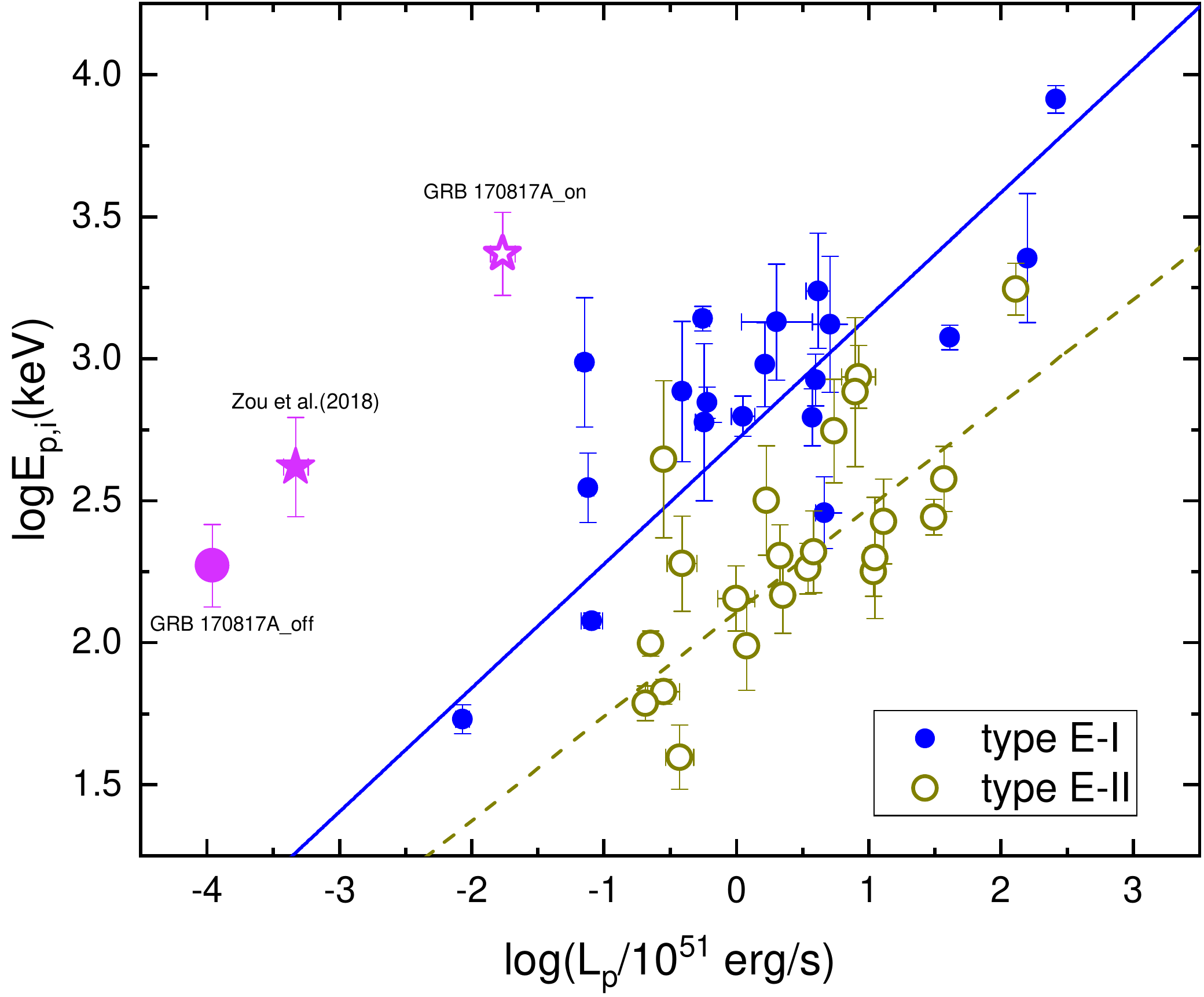}
\caption{The energy relations of $E_{p,i}$ versus $L_{p}$ in logarithmic scale for short and long GRBs in upper panel and for E-I and E-II GRBs in lower panel. All symbols are same as in Figure \ref{sec:fig5}. The straight lines stand for the best fits to data.\\
(A color version of this figure is available in the online journal)}
\label{sec:fig6}
\end{center}
\end{minipage}
\end{figure}

Similarly, we try to fit the Yonetoku relations $E_{p,i}\sim C_2 L_{p}^{\eta_{2}}$ of the above four kinds of EE GRBs in Figure \ref{sec:fig6} and their corresponding formulas are written as
\begin{equation}
 E_{p,i}=464.13^{+82.9}_{-70.3}\left(\frac{L_p}{10^{51}erg/s}\right)^{0.42\pm0.05 } (keV) ,\label{equation:5}
\end{equation}
\begin{equation}
 E_{p,i}=138.11^{+20.7}_{-18.0}\left(\frac{L_p}{10^{51}erg/s}\right)^{0.37\pm0.07 } (keV) ,\label{equation:6}
\end{equation}
\begin{equation}
 E_{p,i}=516.00^{+86.2}_{-73.9}\left(\frac{L_p}{10^{51}erg/s}\right)^{0.44\pm0.05 } (keV) , \label{equation:7}
\end{equation}
\begin{equation}
 E_{p,i}=128.05^{+13.7}_{-12.4}\left(\frac{L_p}{10^{51}erg/s}\right)^{0.37\pm0.05 } (keV) \label{equation:8}
\end{equation}
for short, long, E-I and E-II EE GRBs, respectively. We are aware that the power-law indexes of the four kinds of bursts are approximately consistent with each other as shown in Table \ref{tab2} and they are slightly less than 0.5 that demonstrates the synchrotron radiation to be dominant for the GRBs with EE \citep[see also][]{Zhang+2012,Zhang+2018}. Our results are roughly in agreement with some previous ones \citep{Wei+2003,Yonetoku+2004,Wang+2011,Zhang+2012,Zhang+2018}. No matter whether the off-axis viewing effect is corrected, GRB 170817A is undoubtedly a violator of the Yonetoku relation as illustrated in Figure \ref{equation:6}. After taking into account the same off-axis parameters, one can obtain the on-axis peak luminosity of GRB 170817A to be $L_{p,on}=(1.68\pm0.36)\times10^{49}$ erg/s dimmer than most GRBs.

\begin{table}
\bc
\begin{minipage}[]{100mm}
\caption[]{Parameters of energy correlations for the EE GRBs\label{tab2}}
\end{minipage}
\setlength{\tabcolsep}{12pt}
\small
 \begin{tabular}{ccccccc}
  \hline\noalign{\smallskip}
Type& Correlation &C(keV) &$\eta$ &$ R^b$ & $z_p$ & $A_{z_p} $   \\
  \hline\noalign{\smallskip}
SGRB (N=21$^{a}$)&$ E_{p,i}-E_{iso}$   & $1783.61^{+527.5}_{-407.3}$ & $0.43\pm0.06$  & 0.73  & $ 2.7 $  & $2.98\times10^{12}$ \\
SGRB (N=21$^{a}$)&$ E_{p,i}-L_p  $     & $464.13^{+82.9}_{-70.3}$    & $0.42\pm0.05$  & 0.74  & $ 2.6 $  & $1.61\times10^{12}$ \\ \hline
LGRB (N=20)&$ E_{p,i}-E_{iso}$   & $212.82^{+32.4}_{-28.2}$   & $0.37\pm0.06$  & 0.66  & $ 2.0 $  & $1.19\times10^{11}$ \\
LGRB (N=20)&$ E_{p,i}-L_p  $     & $138.11^{+20.7}_{-18.0}$   & $0.37\pm0.07$  & 0.62  & $ 2.0 $  & $3.00\times10^{11}$ \\ \noalign{\smallskip}\hline
E-I (N=19$^{a}$)&$ E_{p,i}-E_{iso}$   & $2062.76^{+552.9}_{-436.0}$ & $0.45\pm0.05$  & 0.80  & $ 3.0 $  & $2.18\times10^{12}$ \\
E-I (N=19$^{a}$)&$ E_{p,i}-L_p  $     & $516.00^{+86.2}_{-73.9}$    & $0.44\pm0.05$  & 0.80  & $ 2.9 $  & $1.29\times10^{12}$ \\ \noalign{\smallskip}\hline
E-II (N=22)&$ E_{p,i}-E_{iso}$   & $207.60^{+23.5}_{-21.1}$  & $0.36\pm0.04$  & 0.75  & $ 1.9 $  & $1.31\times10^{11}$ \\
E-II (N=22)&$ E_{p,i}-L_p  $     & $128.05^{+13.7}_{-12.4}$  & $0.37\pm0.05$  & 0.73  & $ 2.0 $  & $2.58\times10^{11}$ \\
  \noalign{\smallskip}\hline
\end{tabular}
\ec
\tablecomments{1\textwidth}{$^a$ The short and off-axis GRB 170817A/GW 170817 has not been utilized during our fits.\ \ \ \ \ $^b$ R index is the linear correlation coefficient of these energy relations in logarithmic scale.}

\end{table}

\subsection{Classifying GRBs with energy correlations}\label{sec:result--3}
\begin{figure}
\begin{minipage}[t]{1\linewidth}
\begin{center}
\includegraphics[height=10cm,width=12.0cm]{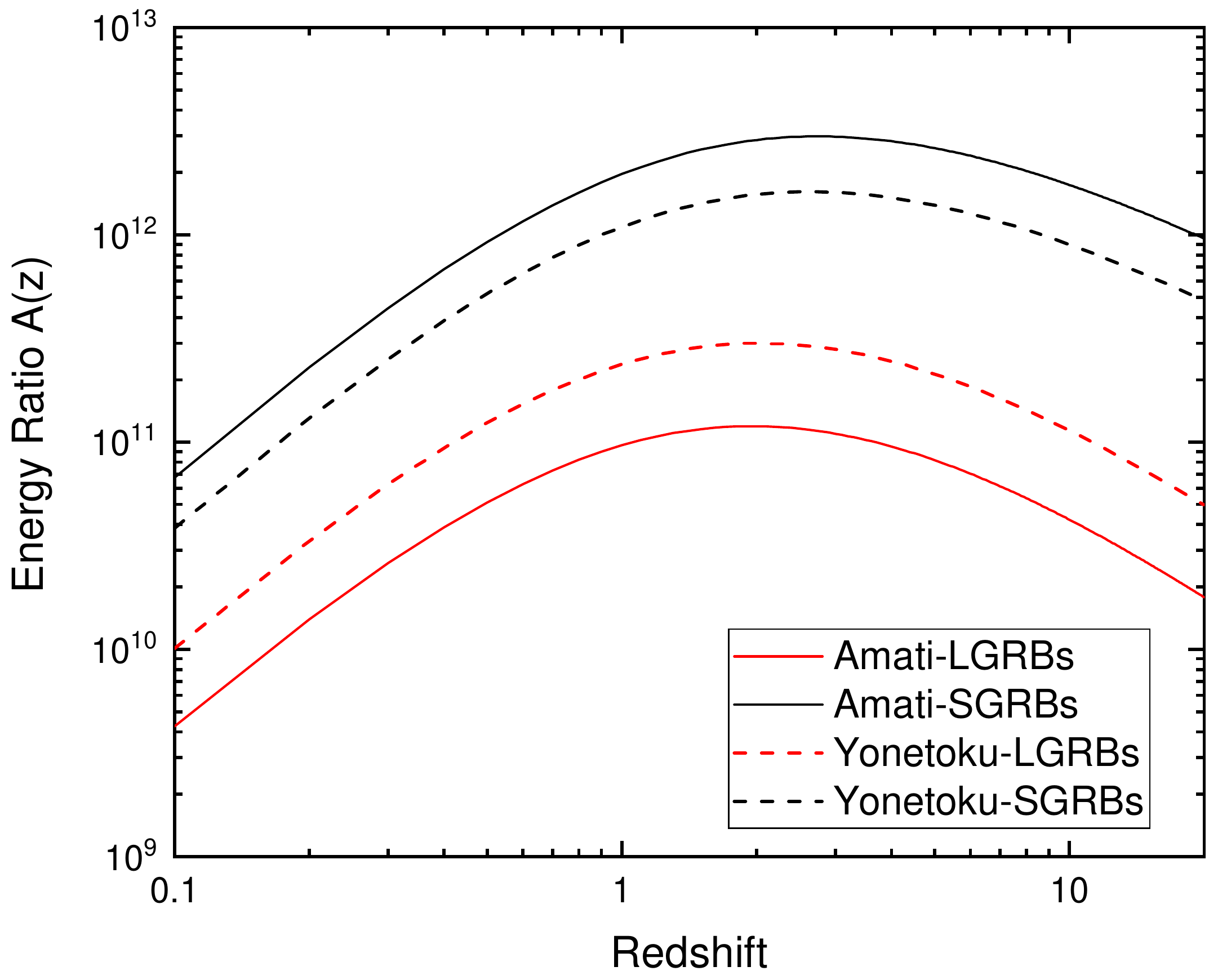}
\end{center}
\end{minipage}
\begin{minipage}[t]{1\linewidth}
\begin{center}
\includegraphics[height=10cm,width=12.0cm]{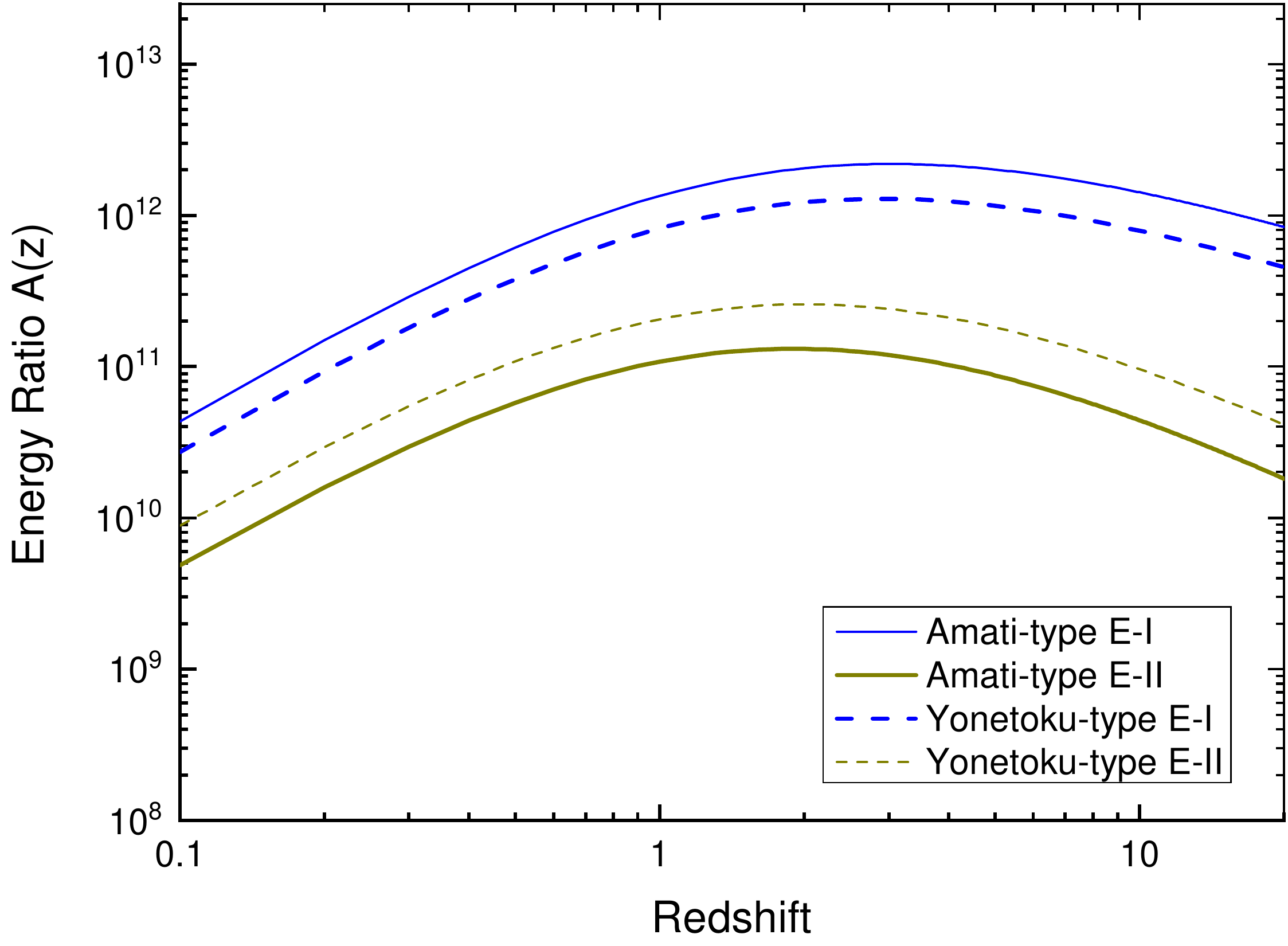}
\caption{The dependencies of $A(z)$ on redshift are gotten for the Amati (solid lines) and Yonetoku (dashed lines) relations of short and long GRBs in the upper panel, and the same two energy relations of E-I and E-II GRBs in the lower panel.\\
   (A color version of this figure is available in the online journal)}
\label{sec:fig7}
\end{center}
\end{minipage}
\end{figure}

\begin{figure}
\begin{minipage}[t]{1\linewidth}
\begin{center}
\includegraphics[height=10cm,width=12.0cm]{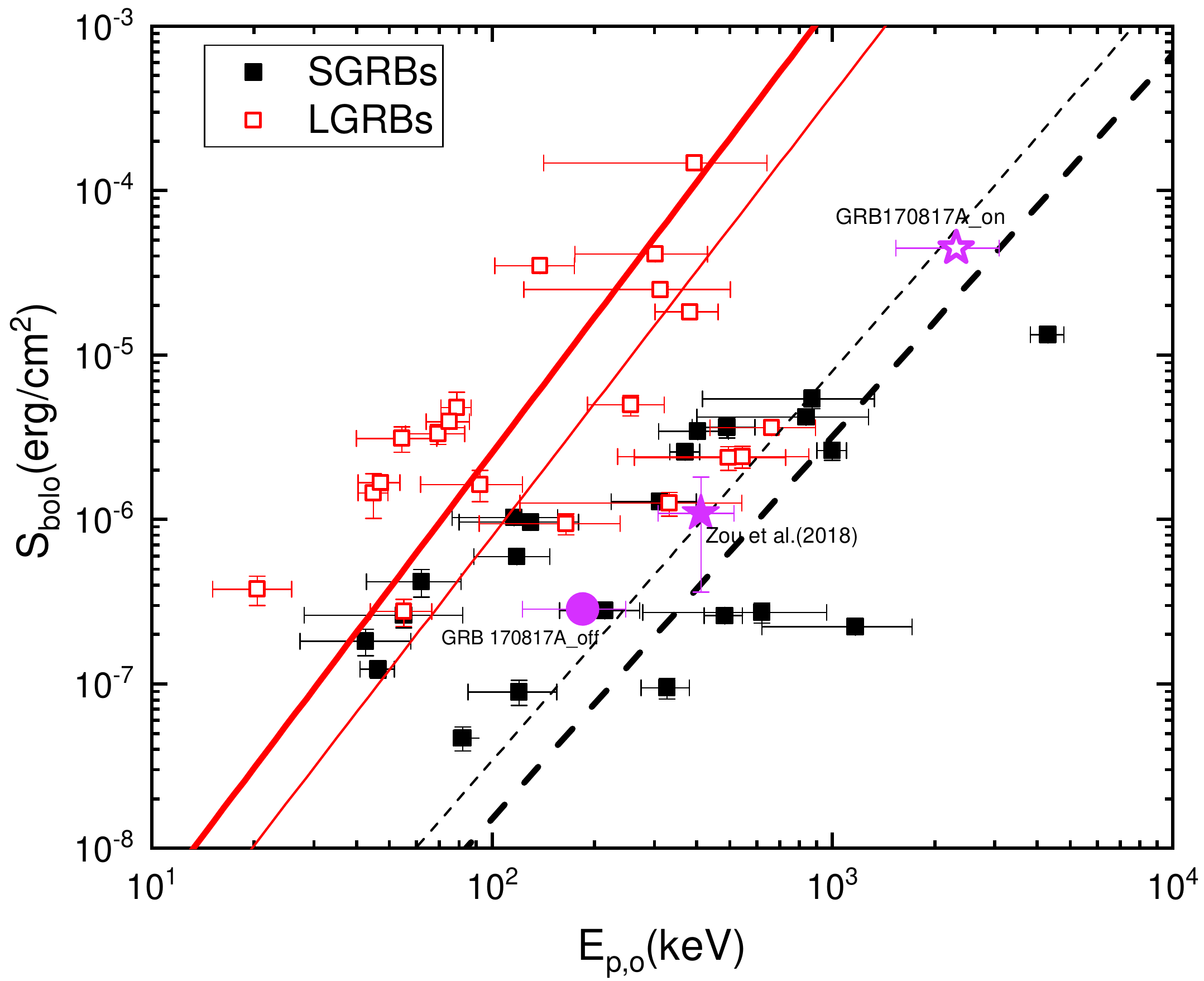}
\end{center}
\end{minipage}
\begin{minipage}[t]{1\linewidth}
\begin{center}
\includegraphics[height=10cm,width=12.0cm]{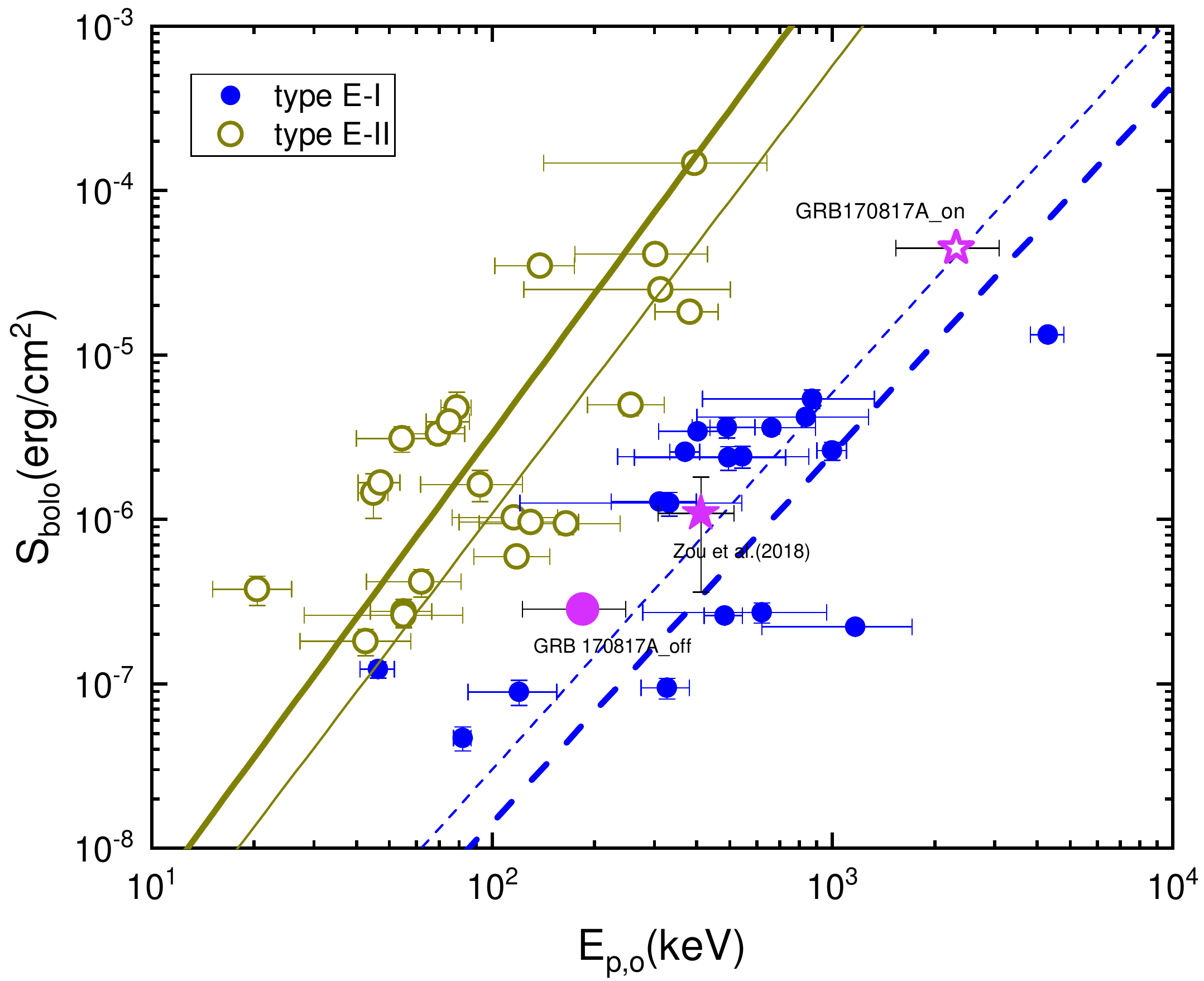}
\caption{The $S_{bolo}$ is plotted against the $E_{p,o}$ for different types of bursts. The lower limits on the $S_{bolo}$-$E_{p,o}$ relationship from the Amati/Yonetoku relations of short (thick/thin black dashed line) versus long (thick/thin red solid line) GRBs and those of E-I (thick/thin blue dashed line) versus E-II (thick/thin dark yellow solid line) bursts are individually presented in the upper and lower panels. All symbols are same as in Figure \ref{sec:fig5}. \\   (A color version of this figure is available in the online journal)}
\label{sec:fig8}
\end{center}
\end{minipage}
\end{figure}

We now apply our new energy correlations of Eqs. \ref{equation:1}-\ref{equation:8} to verify if they can distinguish different kinds of GRBs in the plane of $E_{p,o}$ versus $S_{bolo}$. If substituting $E_{iso}=4\pi D_{l}^{2}S_{bolo}(1+z)^{-1}$ (or $L_p=4\pi D_{l}^{2}P_{bolo}$) into $E_{p,i}=C_1 (E_{iso}/10^{52} erg)^{\eta_{1}}$ (or $E_{p,i}=C_2 (L_{p}/10^{51} erg/s)^{\eta_{2}}$ ) and carrying out variable separations, one can get the energy ratios $\zeta_j=E_{p,o}^{1/\eta_j}/S_{bolo}\propto A_{j}(z) (j=1,2)$ evolving with redshift as shown in Figure \ref{sec:fig7}, in which $A_j(z)$ will reach its maximum values at a certain redshift of $z_p$ (see also paper I for a detail). Table \ref{tab2} lists the values of $z_p$ and $A(z_p)$ constrained with the fitted parameters $C_j$ and $\eta_j$ from Eq. \ref{equation:1} to Eq. \ref{equation:8}. This in return puts a lower limit to the logarithmic relationships of $logS_{bolo}\geq logE_{p,o}/\eta_j-logA(z_p)$ as displayed in Figure \ref{sec:fig8}, where we find that both Amati and Yonetoku relations can be used to classify these EE GRBs themselves, which is very similar to the findings for the whole GRB samples in paper I. To draw the lower limit lines from the Yonetoku relations, $P_{bolo}=S_{bolo}(P_{\gamma}/S_{\gamma})\simeq S_{bolo}/T_{90}$ with a typical duration $T_{90}=2$ s has been applied. Previously, \cite{Qin+2013} also proposed that GRBs are better to be assorted into Amati and non-Amati classes. Note that the non-Amati bursts in \cite{Qin+2013} actually correspond to the SGRBs. Interestingly, these empirical energy correlations are available to identify not only short vs. long but also E-I vs. E-II GRBs. By contrast, the E-I/II classification scheme is more reasonable since the two kinds of bursts are less overlapped. Although GRB 170817A does match neither Amati nor Yonetoku relations, we needs to emphasize that GRB 170817A is always located near to the region of either the short or the E-I GRBs regardless of whether the off-axis effect is considered or not.

\subsection{Spectral hardness}\label{sec:result--4}
\begin{figure}
\begin{minipage}[t]{1\linewidth}
\begin{center}
\includegraphics[height=10cm,width=12.0cm]{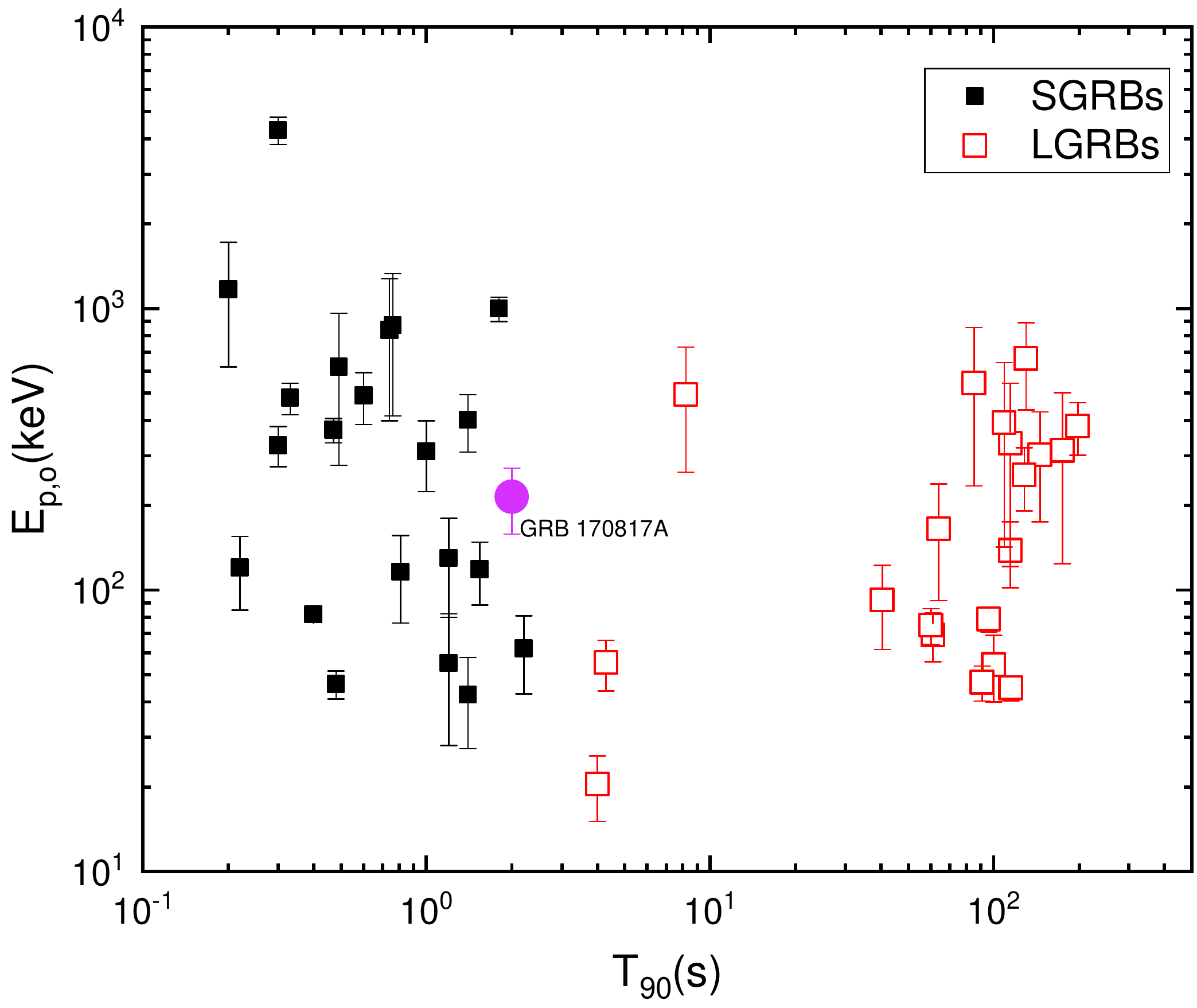}
\end{center}
\end{minipage}
\begin{minipage}[t]{1\linewidth}
\begin{center}
\includegraphics[height=10cm,width=12.0cm]{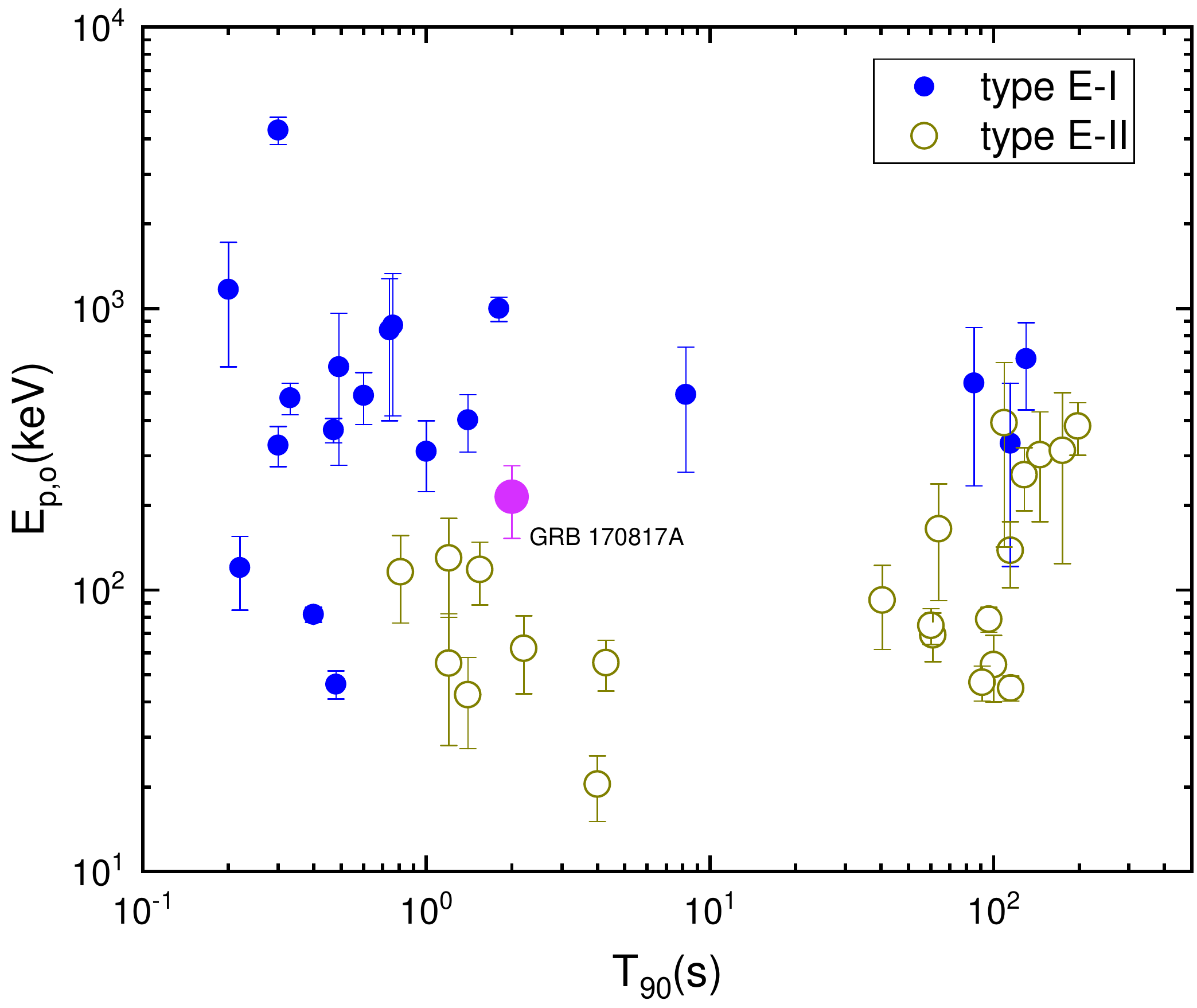}
\caption{Correlations between $T_{90}$ and $E_{p,o}$ for short and long GRBs in the upper panel and for E-I and E-II GRBs in the lower panel. All symbols are same as in Figure \ref{sec:fig5}. GRB 170817A viewed off-axis is marked with larger filled circle. \\
   (A color version of this figure is available in the online journal)}
\label{sec:fig9}
\end{center}
\end{minipage}
\end{figure}

As shown in Figure \ref{sec:fig9}, the $E_{p,o}$ and the $T_{90}$ are weakly anti-correlated with a Pearson correlation coefficient of $\rho =-0.16$ and a chance possibility of 0.4. Interestingly, the E-II bursts tend to have longer $T_{90}$ but smaller $E_{p,o}$ in contrast with the E-I GRBs and both of them exhibit a wider $T_{90}$ span from 0.1 to 200 seconds. On the contrary, the $E_{p,o}$ does not show an obvious dependence on the $T_{90}$ from short to long bursts, which is consistent with some results of BATSE and Swift normal GRBs \citep[e.g.][]{Ghirlanda04,Zhang2020}. It happens that GRB 170817A just lies on the boundaries between short/E-I and long/E-II GRBs, which makes it more mysterious on the aspects of classification.

\textbf{\subsection{Properties of the EE components}}\label{sec:result--5}
In case of the EE component, it also contains many useful parameters that can be utilized to unveil the associations of central engine with the EE formation mechanisms including energy injection effect \citep{Yu+2013,Xu+15}. In this section, we will focus on the comparative studies of the time delay, peak brightness and peak luminosity of the EE segments for 10 short and 19 long GRBs (see those bursts marked with star in Table \ref{tab2}) with well-determined EEs at a higher confidence level of $S/N>3$. Coincidentally, there are 10 E-I and 19 E-II bursts in the selected sub-sample. In addition, the energy correlations of the EE parts will be also investigated to explore the possible connections with the GRB counterparts.

\begin{figure}
\begin{minipage}[t]{1\linewidth}
\begin{center}
\includegraphics[height=10cm,width=12.0cm]{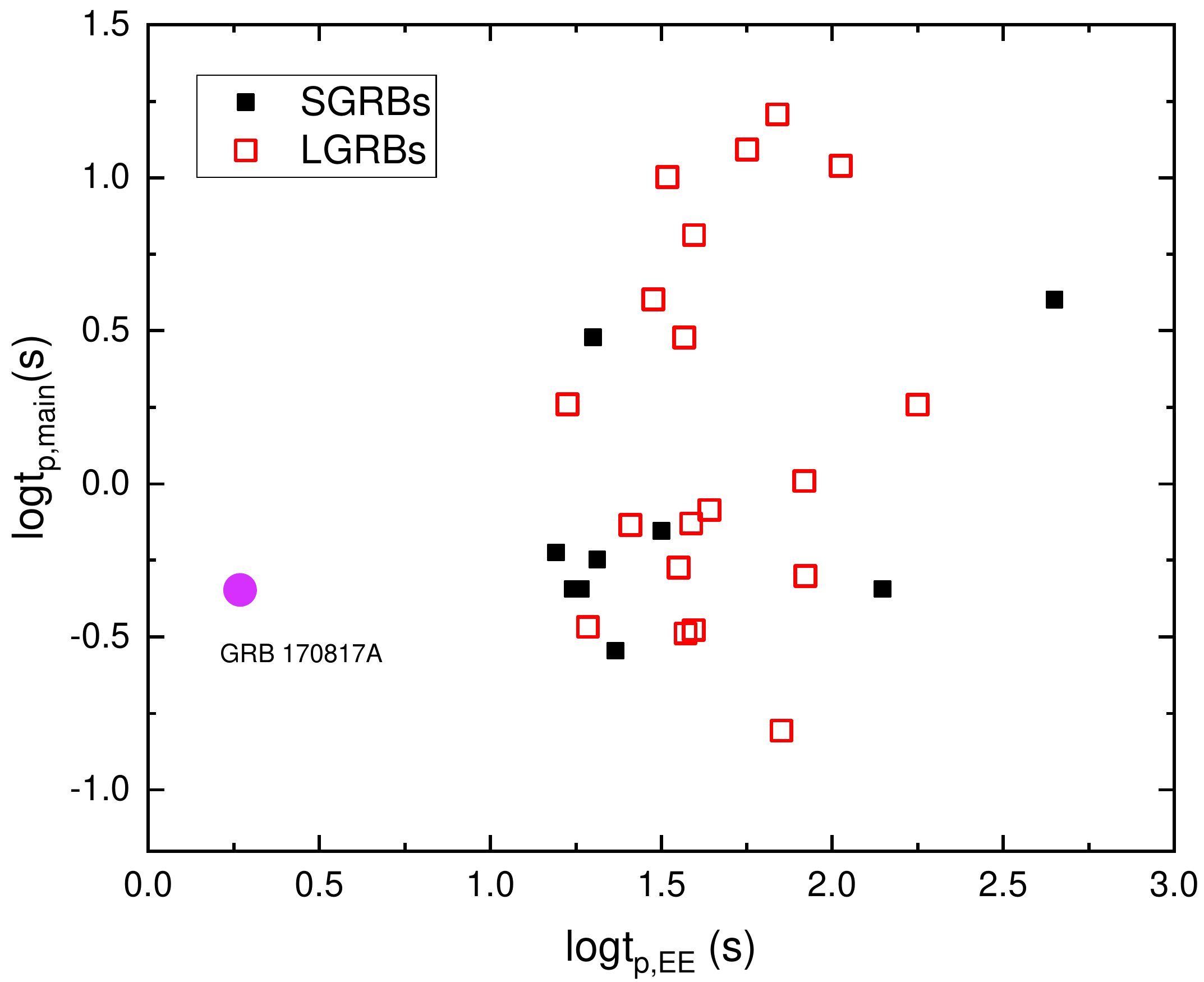}
\end{center}
\end{minipage}
\begin{minipage}[t]{1\linewidth}
\begin{center}
\includegraphics[height=10cm,width=12.0cm]{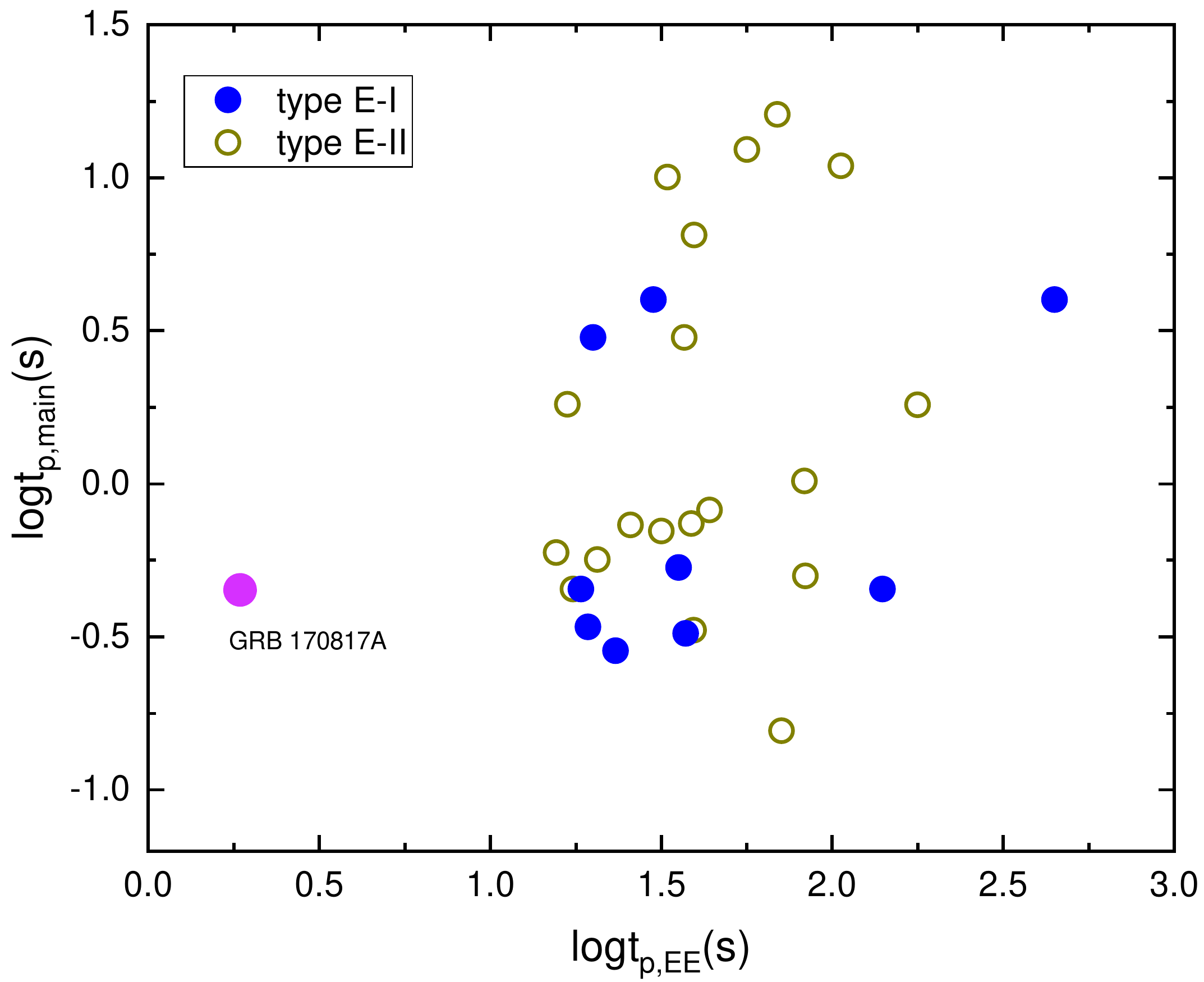}
\caption{Correlations between the peak time of main bursts ($t_{p,main}$) and that of the EE components ($t_{p,EE}$) in logarithmic scale for short (filled square) and long (empty square) GRBs in upper panel and for E-I (filled circle) and E-II (empty circle) GRBs in lower panel. GRB 170817A viewed off-axis is marked with larger filled circle. \\   (A color version of this figure is available in the online journal)}
\label{sec:fig10}
\end{center}
\end{minipage}
\end{figure}

\begin{figure}
\begin{minipage}[t]{1\linewidth}
\begin{center}
\includegraphics[height=10cm,width=12.0cm]{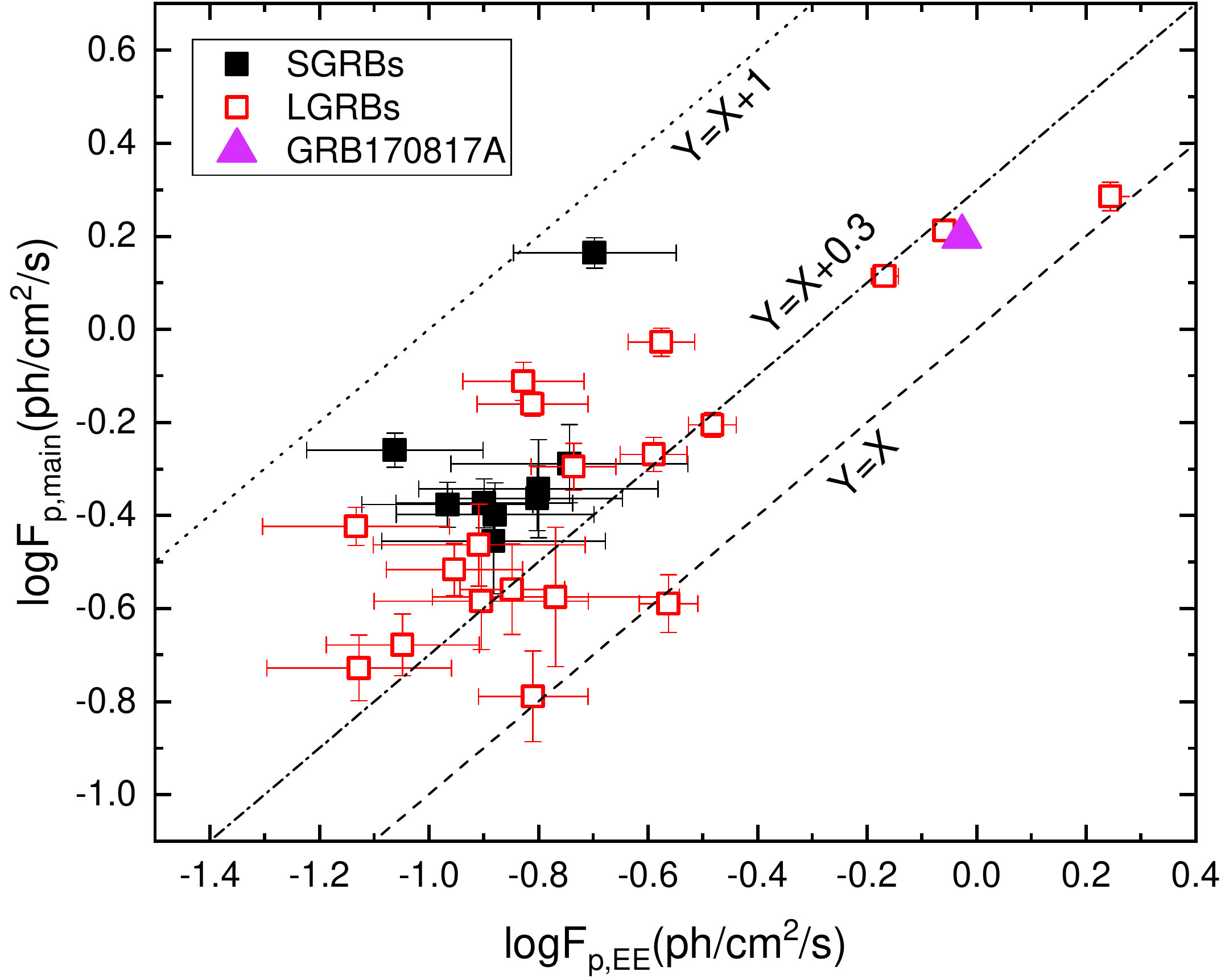}
\end{center}
\end{minipage}
\begin{minipage}[t]{1\linewidth}
\begin{center}
\includegraphics[height=10cm,width=12.0cm]{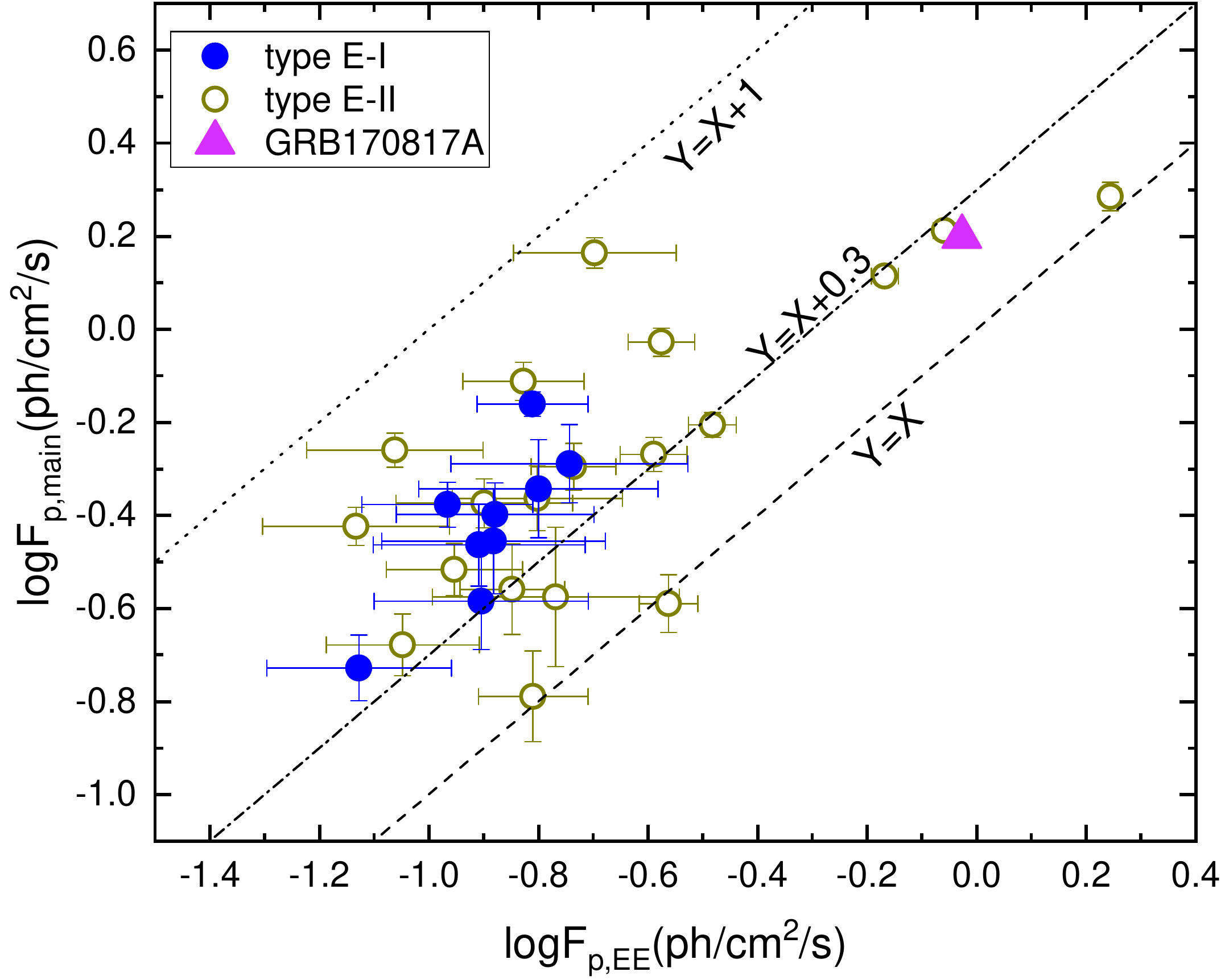}
\caption{Correlations of the peak flux of main bursts ($F_{p,main}$) with that of the EE components ($F_{p,EE}$) in logarithmic scale. All symbols are same as in Figure \ref{sec:fig10}. GRB 170817A viewed off-axis is marked with larger filled triangle. Three peak flux ratios of $F_{p,EE}$ to $F_{p,main}$ are denoted by the dashed, dash-dotted and dotted lines for 1, 1/2 and 1/10, respectively. \\   (A color version of this figure is available in the online journal)}
\label{sec:fig11}
\end{center}
\end{minipage}
\end{figure}

\begin{figure}
\begin{minipage}[t]{1\linewidth}
\begin{center}
\includegraphics[height=10cm,width=12.0cm]{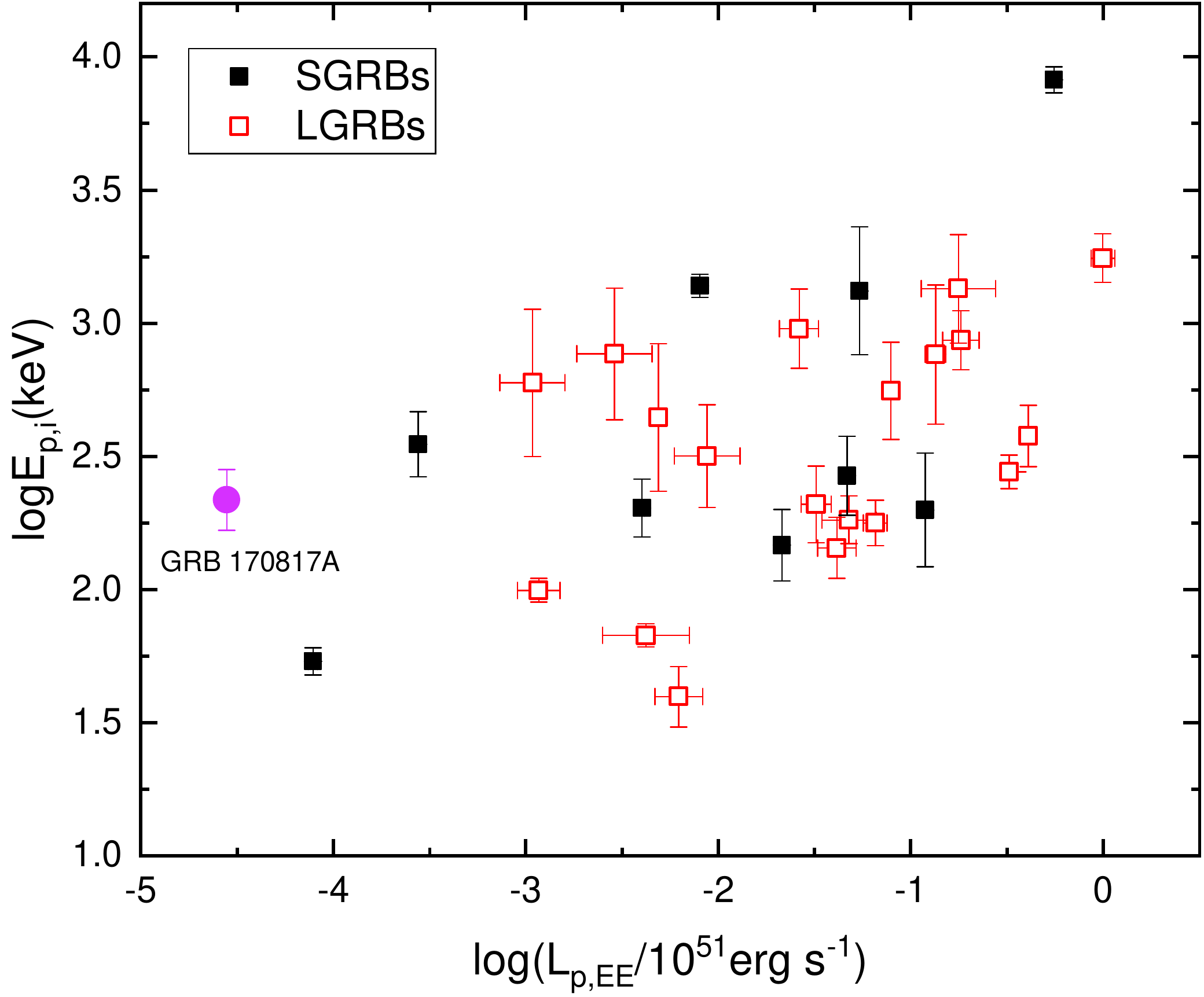}
\end{center}
\end{minipage}
\begin{minipage}[t]{1\linewidth}
\begin{center}
\includegraphics[height=10cm,width=12.0cm]{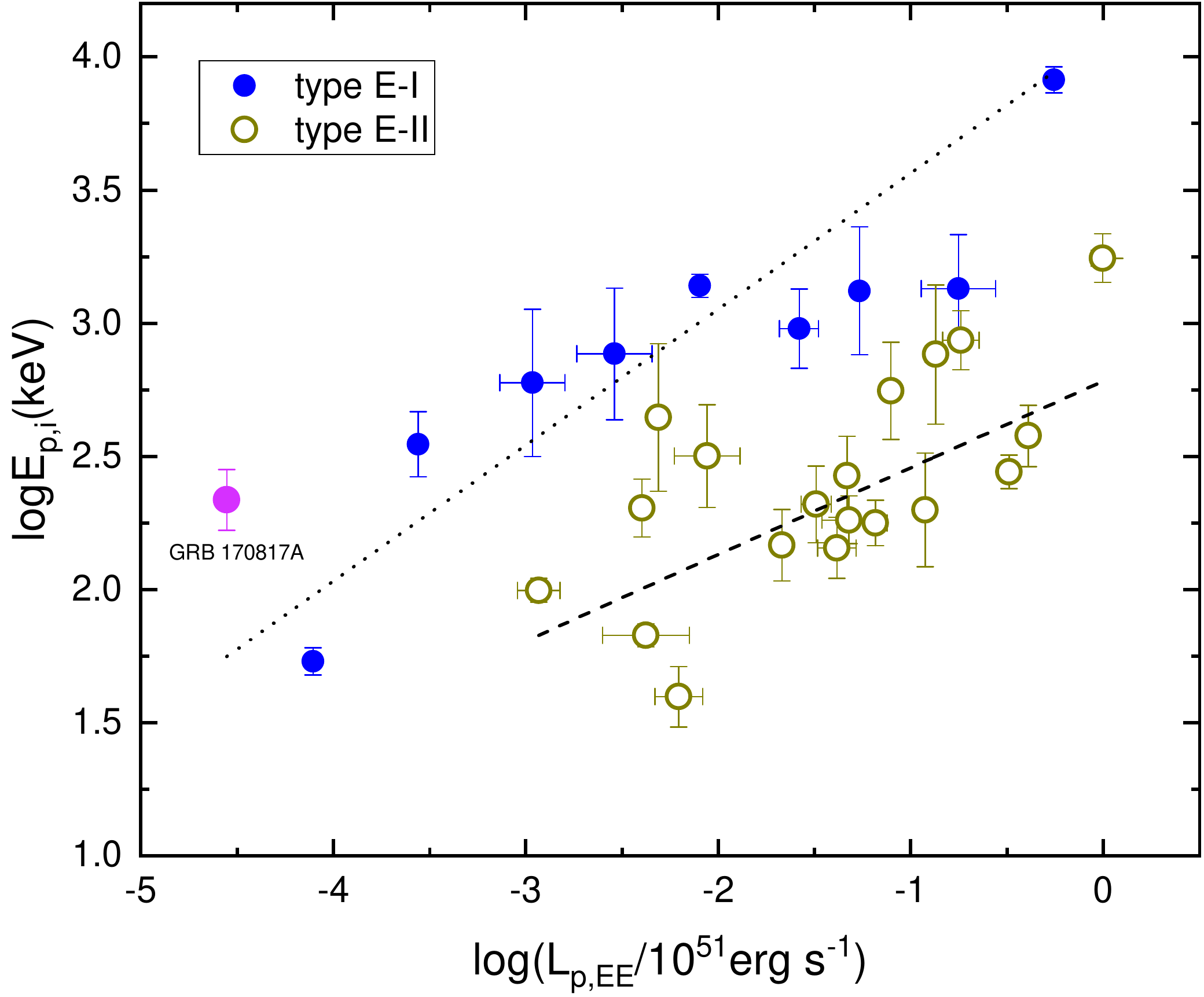}
\caption{The energy correlations of the $E_{p,i}$ with the $L_{p,EE}$ in logarithmic scale. All symbols are same as in Figure \ref{sec:fig10}. GRB 170817A viewed off-axis is marked with larger filled circle. The dashed and the dotted lines represent the best weighted fits to E-I and E-II GRBs in each with a power-law form (see text for details). \\   (A color version of this figure is available in the online journal) }
\label{sec:fig12}
\end{center}
\end{minipage}
\end{figure}

Figure \ref{sec:fig10} indicates that there are no any correlations between the peak time of main bursts ($t_{p,main}$) and the peak time of the EE components ($t_{p,EE}$). Except GRB 170817A with extremely early EE, the majority of GRBs have the EE profiles peaking at a delay time of $45.7_{-20.6}^{+37.5}$ seconds after the trigger. We examine the associations of the peak fluxs of the EE components ($F_{p,EE}$) with those of the main bursts ($F_{p,main}$) in Figure\ref{sec:fig11}, from which one gains the logarithmic correlation coefficients $\rho=$0.48, 0.82, 0.81 and 0.72 with \textit{p}-values of 0.19, $2.1\times10^{-5}$, $7.3\times10^{-3}$ and $5.7\times10^{-4}$ for short, long, E-I and E-II GRBs, correspondingly. These correlations imply that the EE energy outputs should depend on the energy amount of their own main bursts. There are three bursts (GRB 060614, 070223 and 100814A) with stronger EEs comparable to their main peaks. It also can be seen from Figure \ref{sec:fig11} that a large fraction of the EE GRBs have the peak flux ratios of $F_{p,EE}/F_{p,main}$ ranging from 1/10 to 1/2. Figure \ref{sec:fig12} is plotted to test whether the popular Yonetoku relation is existent during the EE phase. To do this, the EE peak luminosity is estimated by $L_{p,EE}=4\pi D_{l}^2F_{p,EE}$. The average $L_{p,EE}$ values are $\sim1.1\times10^{49}$ erg/s with a spread of 2.28 dex and $\sim3.1\times10^{49}$ erg/s with a spread of 1.76 dex for short and long GRBs respectively. While the average $L_{p,EE}$ values are $\sim7.6\times10^{48}$ erg/s with a spread of 2.58 dex and $\sim3.7\times10^{49}$ erg/s with a spread of 1.58 dex for E-I and E-II GRBs, respectively. We can find that the $L_{p,EE}$ is positively correlated with the $E_{p,i}$ for all kinds of EE bursts, especially for the E-I/II bursts. Interestingly, the Yonetoku relations of E-I and E-II GRBs can be individually described by $E_{p,i}\propto L_{p,EE}^{0.51\pm0.05}$ and $E_{p,i}\propto L_{p,EE}^{0.33\pm0.06}$ that are good in agreement with Eqs. \ref{equation:7} and \ref{equation:8} respectively. This confirms again that the EE components should be physically associated with the prompt GRBs (See also \citealt{Li+2020b}).

\section{conclusion}\label{sec:Conclusion}
We have selected a complete sample of GRBs with extended emission to study their parameter properties and the possible connections between the softer EE components and the harder GRBs. Simultaneously, we checked whether some previous energy correlations still hold for these particular bursts and how to use the newly-built energy correlations to classify this kind of EE bursts. Our major findings of this work are summarized below:
\begin{itemize}
\item[*] Unlike short and long GRBs, the redshift distributions of E-I and E-II bursts are found to take from different parent distribution. The redshift distributions of E-I and E-II GRBs are different and their medians are 0.51 and 1.29, respectively.
\item[*] The spectral peak energies in the observer and the source frames are identically distributed for short and long bursts but differently distributed for E-I and E-II GRBs, of which their mean $E_{p,o}$ values are correspondingly $\sim$422.7 keV and $\sim$97.7 keV.
\item[*] We find that the Amati and Yonetoku energy correlations do exist for not only short and long EE bursts but also E-I and E-II GRBs. By contrast, the $E_{p,i}$-$E_{iso}$ and $E_{p,i}$-$L_{p}$ power-law relations of E-I and E-II bursts are more tight than those of short and long ones. In addition, the power-law indexes of these energy relations are marginally consistent with most previous values of normal GRBs.
\item[*] Particularly, we notice that GRB 170817A as the first gravitational-wave associated SGRB with EE does not obey either the Amati or the Yonetoku relations no matter whether it was viewed off-axis or on-axis. However, GRB 170817A is located among the region of short or E-I bursts in the $E_{p,o}$ versus $S_{bolo}$ plot.
\item[*] It is confirmed again that the EE GRBs can be identified by the diverse Amati and Yonetoku correlations in the $E_{p,o}$-$S_{bolo}$ plane, which is similar to the conclusion in paper I for ordinary GRBs. Furthermore, E-I and E-II GRBs can be clearly distinguished according their different Amati or Yonetoku energy relations, redshift distributions and peak energy distributions, which demonstrates that the classification scheme of E-I and E-II bursts is more reasonable.
\item[*] Most EE segments in our sample are found to peak at a time of $45.7_{-20.6}^{+37.5}$s after trigger occurrence that is not related with the peak time of main bursts. However, peak fluxes of the EE components and the GRBs are strongly correlated with each other. Surprisingly, we find that the $E_{p,i}$ of GRBs and the EE peak luminosity of E-I/II bursts are also tightly connected with the coincident power-law indexes as those fitted by the normal GRBs only.

\end{itemize}

On the basis of these comparative studies, we conclude that it is much better to reclassify the bursts with EE into two subgroups, that is type E-I  and type E-II, respectively. Therefore, we hope that the most important role of our results could show new lights on the physics of the EE GRBs together with their mysterious progenitors, especially on how to classify or find more EE GRBs resembling the attractive but challenging GRB 170817A.

\normalem
\begin{acknowledgements}
We appreciate the referee for the constructive suggestion and comments that makes the paper improved greatly. We thank Y. Zhang and K. Zhang for their helpful discussions. This work was supported by the Youth Innovations and Talents Project of Shandong Provincial Colleges and Universities (Grant No. 201909118) and the Natural Science Foundations (ZR2018MA030, XKJJC201901 and OP201511). HYC was supported by a National Research Foundation of Korea Grant funded by the Korean government (NRF2018R1D1A3B070421880 and 2018R1A6A1A06024970).

\end{acknowledgements}

\bibliographystyle{raa}

\begin{thebibliography}{99}


\bibitem[Ahlgren et al.(2019)]{Ahlgren+2019} Ahlgren, B., Larsson, J., Valan, V., et al.\ 2019, \apj, 880, 76

\bibitem[Amati et al.(2002)]{Amati+2002} Amati, L., et al.\ 2002, \aap, 390, 81

\bibitem[Amati (2006)]{Amati+2006} Amati, L.\ 2006, \mnras, 372, 233

\bibitem[Amati (2012)]{Amati+2012} Amati, L. 2012, IJMPS, 12, 19

\bibitem[Amati et al.(2019)]{Amati+2019} Amati, L., et al., 2019, \mnras, 486, L46



\bibitem[Barthelmy et al.(2005)] {Barthelmy+05}Barthelmy, S. D., Cannizzo, J. K., Gehrels, N., et al., 2005. \apjl, 635, L33


\bibitem[Bostanci et al.(2013)]{Bostanci+13}Bostanci, Z. F., Kaneko, Yuki; Gogus, Ersin, 2013, \mnras, 428, 1623

\bibitem[Bucciantini et al.(2012)] {Bucciantini+2012} Bucciantini, N., Metzger, B. D., Thompson, T. A., Quataert, E.\ 2012, \mnras, 419, 153

\bibitem[Bulik et al.(1998)]{Bulik+1998} Bulik, T., Belczy¨½ski, K., Zbijewski, W.\ 1999, \mnras, 309, 629

\bibitem[Butler et al.(2007)]{Butler+2007} Butler, N. R., Kocevski, D., Bloom, J. S., et al.\ 2007, \apj, 671, 656


\bibitem[Chattopadhyay et al.(2007)]{Chattopadhyay+2007} Chattopadhyay, T., Misra, R., Chattopadhyay, Asis, K., Naskar, M.\ 2007, \apj, 667, 1017

\bibitem[Chattopadhyay et al.(2019)]{Chattopadhyay+2019} Chattopadhyay, S., Maitra, R.\ 2019, \mnras, 481, 3196

\bibitem[Connaughton (2002)]{Connaughton+02}Connaughton, V., 2002, \apj, 567, 1028

\bibitem[Dainotti et al (2018)]{Dainotti+2018} Dainotti, M. G.; Amati, L.\ 2018, \pasp, 130, 051001


\bibitem[Fan et al.(2006)]{Fan+2006} Fan, Y. Z., \& Xu, D.\ 2006, \mnras, 372, L19




\bibitem[Fruchter et al.(2006)]{Fruchter+2006} Fruchter, A.~S., Levan, A.~J., Strolger, L., et al.\ 2006, \nat, 441, 463

\bibitem[Fryer et al.(1999)]{Fryer+1999} Fryer, C. L., Woosley, S. E., Hartmann, D. H.\ 1999, \apj, 526, 152

\bibitem[Galama et al.(1998)]{Galama+1998} Galama, T. J., Vreeswijk, P. M., van Paradijs, J., et al.\ 1998, \ Nature, 395, 670

\bibitem[Gehrels et al.(2004)]{Ghirland+2004} Gehrels, N., Chincarini, G., Giommi, P., et al.\ 2004, \apj, 611, 1005



\bibitem[Ghirlanda et al.(2004)]{Ghirlanda04} Ghirlanda, G., Ghisellini, G., \& Celotti, A., 2004, \aap, 422, L55


\bibitem[Gibson et al.(2017)]{Gibson+2017} Gibson, S. L., Wynn, G. A., Gompertz, B. P., et al.\ 2017, \mnras, 470, 4925

\bibitem[Goldstein al.(2017)]{Goldstein+17}Goldstein, A., Veres, P., Burns, E., et al. 2017, \apjl, 848, L14






\bibitem[Gompertz et al.(2013)]{Gompertz+2013} Gompertz, B.~P., O'Brien, P.~T., Wynn, G.~A., et al.\ 2013, \mnras, 431, 1745

\bibitem[Gompertz et al.(2014)]{Gompertz+2014} Gompertz, B.~P., O'Brien, P.~T., \& Wynn, G.~A.\ 2014, \mnras, 438, 240

\bibitem[Gompertz et al.(2020)]{Gompertz+2020} Gompertz, B.~P., Levan, A.~J., \& Tanvir, N.~R.\ 2020, \apj, 895, 58

\bibitem[Hajela et al.(2019)]{Hajela+19}Hajela, A., Margutti, R., Alexander, K. D., et al., 2019, \apj, 886, L17


\bibitem[Hjorth et al.(2003)]{Hjorth+2003} Hjorth, J., Sollerman, J., M{\o}ller, P., et al.\ 2003, \nat, 423, 847

\bibitem[Horv{\'a}th, \& T{\'o}th(2016)]{Horvath+2016} Horv{\'a}th, I., \& T{\'o}th, B.~G.\ 2016, \apss, 361, 155

\bibitem[Ioka et al.(2005)]{Ioka+05}Ioka, K., Kobayashi, S., Zhang, B., 2005, \apj, 631, 429

\bibitem[Kaneko et al.(2015)]{Kaneko+2015} Kaneko, Y., Bostanc{\i}, Z.~F., G{\"o}{\u g}{\"u}{\c s}, E., \& Lin, L.\ 2015, \mnras, 452, 82

\bibitem[Kinugawa et al.(2019)]{Kinugawa+2019} Kinugawa, T., Harikane, Y., Asano, K.\ 2019, \apj, 878, 128


\bibitem[Kisaka et al.(2017)]{Kisaka+2017} Kisaka, S., Ioka, K., \& Sakamoto, T.\ 2017, \apj, 846, 142

\bibitem[Klebesadel et al.(1973)]{Klebesadel+1973} Klebesadel, R. W., Strong, I. B., \& Olson, R. A.\ 1973, \apjl, 182, L85

\bibitem[Kouveliotou et al.(1993)]{Kouveliotou+1993} Kouveliotou, C., Meegan, C. A., Fishman, G. J., et al.\ 1993, \apj, 413, 101

\bibitem[Li  et al.(2018)]{Li+2018} Li, B., Li, L. B., Huang, Y. F., et al., 2018, \apjl, 859, L3

\bibitem[Li  et al.(1998)]{Li+1998} Li, L. X., Paczynski, B.\ 1998, \apj, 507, L59

\bibitem[Li et al.(2020a)]{Li+2020a} Li, X. J., Zhang, Z. B., Zhang, C. T., et al., 2020a, \apj, 892, 113

\bibitem[Li et al.(2020b)]{Li+2020b} Li, X. J., Zhang, Z. B., Zhang, X. L., et al., 2020b, \apjs, submitted


\bibitem[Liang \& Zhang (2005)]{Liang+2005} Liang, E. W., Zhang, B.\ 2005, \apj, 633, 611



\bibitem[Mazets et al.(2004)]{Mazets+2004} Mazets, E. P., Aptekar, R. L., et al.\ 2004,\ ASPC, 312, 102

\bibitem[Melandri et al.(2014)]{Melandri+2014} Melandri, A., Pian, E., D'Elia, V., et al.\ 2014, \aap, 567, A29

\bibitem[Metzger et al.(2008)]{Metzger+2008} Metzger, B. D., Quataert, E., Thompson, T. A.\ 2008, \mnras, 385, 1455


\bibitem[Norris et al.(1995)]{Norris+1995} Norris, J. P., Bonnell, J. T., Nemiroff, R. J.\ 1995, \apj,439,542

\bibitem[Norris et al.(2006)]{Norris+2006} Norris, J. P., Bonnell, J. T.\ 2006, \apj, 643, 266

\bibitem[Norris et al.(2000) ]{Norris+2000} Norris, J. P., Marani, G. F., \& Bonnell, J. T.\ 2000, \apj, 534, 248

\bibitem[Paciesas et al.(1999)]{Paciesas+1999} Paciesas, W. S., Meegan, C. A., Pendleton, G. N., et al.\ 1999, \apjs, 122, 465

\bibitem[Popham et al.(1998)]{Popham+1998} Popham, R., Woosley, S. E., Fryer, C.\ 1999, \apj, 518, 356

\bibitem[Qi et al.(2012)]{Qi+2012} Qi, S., \& Lu, T.\ 2012, \apj, 749, 99

\bibitem[Qin \& Chen(2013)]{Qin+2013} Qin, Y. P., Chen, Z. F.\ 2013,\mnras, 430, 163


\bibitem[Reichart et al.(2001)]{Reichart+2001} Reichart, D. E., Lamb, D. Q., Fenimore, E. E., et al.\ 2001, \apj, 552, 57

\bibitem[Salafia et al.(2018)]{Salafia+18}Salafia, O. S., Ghisellini, G., Ghirlanda, G., et al., 2018, \aap, 619, 18


\bibitem[Schaefer et al.(2003)]{Schaefer+2003} Schaefer, B. E.\ 2003, \apj, 583, L71

\bibitem[Schaefer et al.(2007)]{Schaefer+2007} Schaefer, B. E.\ 2007, \apj, 660, 16

\bibitem[Svinkin et al.(2008)]{Svinkin+2008} Svinkin, D. S., Frederiks, D. D., Aptekar, R. L., et al.\ 2016,\ ApJS, 224, 10
\bibitem[Tarnopolski (2019a)]{Tarnopolski+19a}Tarnopolski, M., 2019a, \apj, 887, 97
\bibitem[Tarnopolski (2019b)]{Tarnopolski+19b}Tarnopolski, M., 2019b, \apj, 870, 105


\bibitem[T{\'o}th et al.(2019)]{Toth+2019} T{\'o}th, B.~G., R{\'a}cz, I.~I., \& Horv{\'a}th, I.\ 2019, \mnras, 486, 4823

\bibitem[Troja et al.(2008)]{Troja+2008} Troja, E., King, A.~R., O'Brien, P.~T., et al.\ 2008, \mnras, 385, L10

\bibitem[van Putten et al.(2014)]{van+2014} van Putten, M. H. P. M., Lee, G. M., Della Valle, M., et al. \ 2014, \mnras, 444, L58

\bibitem[Wang et al.(2011)]{Wang+2011} Wang, F. Y., Qi, S., \& Dai, Z. G.\ 2011, \mnras, 415, 3423

\bibitem[Wei \& Gao (2003)]{Wei+2003} Wei, D. M., \& Gao, W. H.\ 2003, \mnras, 345, 743

\bibitem[Wiggins et al.(2018)]{Wiggins+2018} Wiggins, B. K., Fryer, C. L., Smidt, J. M., et al.\ 2018, \apj, 865, 27

\bibitem[Xu \& Huang (2012)]{Xu+2012} Xu, M., Huang, Y. F.\ 2012, \aap, 538, A134
\bibitem[Xu \& Huang (2015)]{Xu+15} Xu, M., Huang, Y. F.\ 2015, \raa, 15, 986

\bibitem[Yonetoku et al.(2004)]{Yonetoku+2004} Yonetoku, D., Murakami, T., Nakamura, T., et al.\ 2004, \apj, 609, 935


\bibitem[Yu \& Huang (2013)]{Yu+2013}Yu Y. B., Huang Y. F., 2013, RAA, 13, 662

\bibitem[Yu et al.(2020)]{Yu+2020} Yu Y. B., Li L. B., Li B., Geng J. J., Huang Y. F., 2020, New Astronomy, 75, 101306

\bibitem[Zhang, \& M{\'e}sz{\'a}ros(2002)]{Zhang+2002} Zhang, B., \& M{\'e}sz{\'a}ros, P.\ 2002, \apj, 581, 1236

\bibitem[Zhang et al.(2014)]{Zhang+2014} Zhang, B. B., Zhang, B., Murase, K., et al.\ 2014, \apj, 787, 66Z

\bibitem[Zhang et al.(2006)]{Zhang+2006} Zhang, Z. B., Deng, J. G., Lu, R. J., et al.\ 2006,\ Chin. J. Astron. Astrophys., 6, 312

\bibitem[Zhang \& Choi (2008)]{Zhang+2008} Zhang, Z. B., \& Choi, C. S.\ 2008, \aap, 484, 293

\bibitem[Zhang et al.(2008)]{Zhang+08} Zhang, Z. B., Xie, G. Z.,\& Choi, C. S.\ 2008,\ Int. J. Mod. Phys. D, 17, 1391


\bibitem[Zhang et al.(2012)]{Zhang+2012} Zhang, Z. B., Chen, D. Y. \& Huang, Y. F.\ 2012, \apj, 755, 55

\bibitem[Zhang et al.(2016)]{Zhang+2016} Zhang, Z. B., Yang, E. B., Choi, C. S., Chang, H. Y. \ 2016, \mnras, 462, 3243

\bibitem[Zhang et al.(2018)]{Zhang+2018} Zhang, Z. B., Zhang, C. T., Zhao, Y. X., et al.\ 2018, \ PASP, 130, 054202 (Paper I)

\bibitem[Zhang et al.(2020)]{Zhang2020}Zhang, Z. B., Jiang, M., Zhang, Y., et al.\ 2020, \apj, submitted

\bibitem[Zitouni et al.(2015)]{Zitouni+2015} Zitouni, H., Guessoum, N., Azzam, W. J., Mochkovitch, R.\ 2015,\ Ap\&SS, 357, 7

\bibitem[Zitouni et al.(2018)]{Zitouni+2018} Zitouni, H., Guessoum, N., AlQassimi, K. M., Alaryani, O.\ 2018,\ Ap\&SS, 363, 223

\bibitem[Zou et al.(2018)]{Zou+18}Zou, Y. C., Wang, F. F., Moharana, R., et al., 2018, \apj, 851, L1

\end{thebibliography}

\end{document}